\documentclass[structabstract]{aa}  
\usepackage{graphicx}
\usepackage{rotating}
\usepackage{natbib}
\usepackage{txfonts}
\begin{document}
   \title{Surface brightness profile of the Milky Way's nuclear star cluster}

   \author{
R. Sch\"odel
\inst{1}
\and
A. Feldmeier 
\inst{2}
\and
D. Kunneriath
\inst{3}
\and
S. Stolovy
\inst{4}
\and
N. Neumayer
\inst{2}
\and
P. Amaro-Seoane\inst{5}
\and
S. Nishiyama
\inst{6}
          }
 \institute{
    Instituto de Astrof\'isica de Andaluc\'ia (CSIC),
     Glorieta de la Astronom\'ia s/n, 18008 Granada, Spain
              \email{rainer@iaa.es}
              \and
    European Southern Observatory, Karl-Schwarzschild-Strasse 2, D-85748 Garching bei M\"unchen, Germany
              \and
   Astronomical Institute, Academy of Sciences, Bo\v{c}n\'i II 1401, 14100 Prague, Czech Republic
             \and
    El Camino College,  16007 Crenshaw Blvd. Torrance, CA 90506, USA
             \and
    Max-Planck Institut f\"ur Gravitationsphysik (Albert-Einstein-Institut), Am M\"uhlenberg 1, D-14476 Potsdam, Germany
             \and
    National Astronomical Observatory of Japan, Mitaka, Tokyo 181-8588, Japan
             }

   \date{Received 21 January 2014 / Accepted 26 March 2014}

 
  \abstract
  {Although the Milky Way nuclear star cluster (MWNSC) was discovered more than four decades ago, several of its key properties have not been determined unambiguously up to now because of the strong and spatially highly variable interstellar extinction toward the Galactic centre.}
  {In this paper we aim at determining the  shape, size, and luminosity/mass of the MWNSC.}
  {To investigate the properties of the MWNSC, we used Spitzer/IRAC images at $3.6$ and $4.5\,\mu$m, where interstellar extinction is at a minimum but the overall emission is still dominated by stars. We corrected the $4.5\,\mu$m image  for PAH emission with the help of the IRAC $8.0\,\mu$m map and for extinction with the help of a $[3.6-4.5]$ colour map. Finally, we investigated the symmetry of the nuclear cluster and fit it with S\'ersic, Moffat, and King models. }
  {We present an extinction map for the central $\sim300\times200$\,pc$^{2}$ of the Milky Way\thanks{The extinction map and the corresponding uncertainty map  shown in Fig.\,\ref{Fig:ext} are made available in electronic form at the CDS via anonymous ftp to cdsarc.u-strasbg.fr (130.79.128.5) or via http://cdsweb.u-strasbg.fr/cgi-bin/qcat?J/A+A/566/A47}, as well as a PAH-emission- and extinction-corrected image of the stellar emission, with a resolution of about $0.20$\,pc. We find that the MWNSC appears in projection to be intrinsically point-symmetric, that it is significantly flattened, with its major axis aligned along the Galactic plane, and that it is centred on the black hole, Sagittarius\,A*. Its density follows the well known approximate $\rho\propto r^{-2}$-law at distances of a few parsecs from Sagittarius\,A*, but becomes as steep as  $\rho\propto r^{-3}$ at projected radii around 5\,pc. We derive a half light radius of $4.2\pm0.4$\,pc,  a total luminosity of $L_{MWNSC,4.5\,\mu{m}}=4.1\pm0.4\times10^{7}\,L_{\odot}$,  and a mass of $M_{MWNSC}=2.5\pm0.4\times10^{7}$\,M$_{\odot}$.}
   {The overall properties of the MWNSC agree well with the ones of its extragalactic counterparts, which underlines its role as a template for these objects. Its flattening agrees well with its previously established rotation parallel to Galactic rotation and suggests that it was formed by accretion of material that tended to fall in along the Galactic plane.  Our findings support the in situ growth scenario for nuclear clusters and emphasise the need to increase the complexity of theoretical models for their formation and for the interaction between their stars and the central black hole in order to include rotation, axisymmetry, and growth in recurrent episodes.}


   \maketitle
%

\section{Introduction}


Nuclear star clusters (NSCs) have been detected in $\sim$75\% of all galaxies and appear as compact clusters at the photometric and dynamical centres of their host galaxies \citep[e.g.,][]{Boker:2002kx,Carollo:1998fk,Cote:2006eu,Neumayer:2011uq}. They have luminosities in the range of $10^{5}-10^{8}$\,L$_{\odot}$, effective radii of a few pc, and masses of $10^{6}-10^{8}$\,M$_{\odot}$. They are typically one to two orders of magnitude brighter and more massive than globular clusters \citep{Boker:2004oq,Walcher:2005ys}, which places NSCs among the most massive known clusters in the Universe \citep[for a brief review, see][]{Boker:2010ys}. Star formation in NSCs appears to be a recurrent process. The majority of NSCs have mixed old and young stellar populations and frequently  show signs of star formation within the past 100\,Ma \citep[e.g.,][]{Rossa:2006zr,Seth:2006uq,Walcher:2006ve}. NSCs show complex morphologies and can coexist with massive black holes \citep[MBHs; ][]{Seth:2008kx,Graham:2009lh,Neumayer:2012fk}.

The study of NSCs can serve to make progress in a variety of astrophysical fields, such as (1) the accretion history of galactic nuclei. While MBHs are the final product of accretion, with only their mass (and perhaps angular momentum) as measurable parameters, NSCs provide a record of the accretion {\it history} through their multiple stellar populations. (2) Since NSCs are, on average, the densest observable stellar systems \citep{Walcher:2005ys,Misgeld:2011kx} and may frequently contain MBHs, they play a key role in the study of stellar dynamics, for example in tests of fundamental ideas such as the formation of stellar cusps around MBHs. Also, phenomena such as tidal disruption events or extreme mass-ratio infall events (so-called EMRIs), which are considered to be important potential sources for gravitational wave emission, are thought to occur in NSCs containing MBHs.  (3) Star formation in NSCs probably proceeds under extreme conditions, at least if we consider the centre of our own Galaxy as representative, which is characterised by a strong Galactic tidal field \citep[e.g.,][]{Portegies-Zwart:2002fk}, high stellar densities \citep[e.g.,][]{Schodel:2007tw}, an intense magnetic field \citep{Crocker:2010fk}, strong UV radiation \citep{Launhardt:2002nx}, and high turbulence and temperature of the interstellar medium \citep[ISM; e.g.,][]{Morris:1996vn}. NSCs can thus help us explore the limits of our understanding of star formation.

\begin{figure*}[!htb]
\includegraphics[width=\textwidth,angle=0]{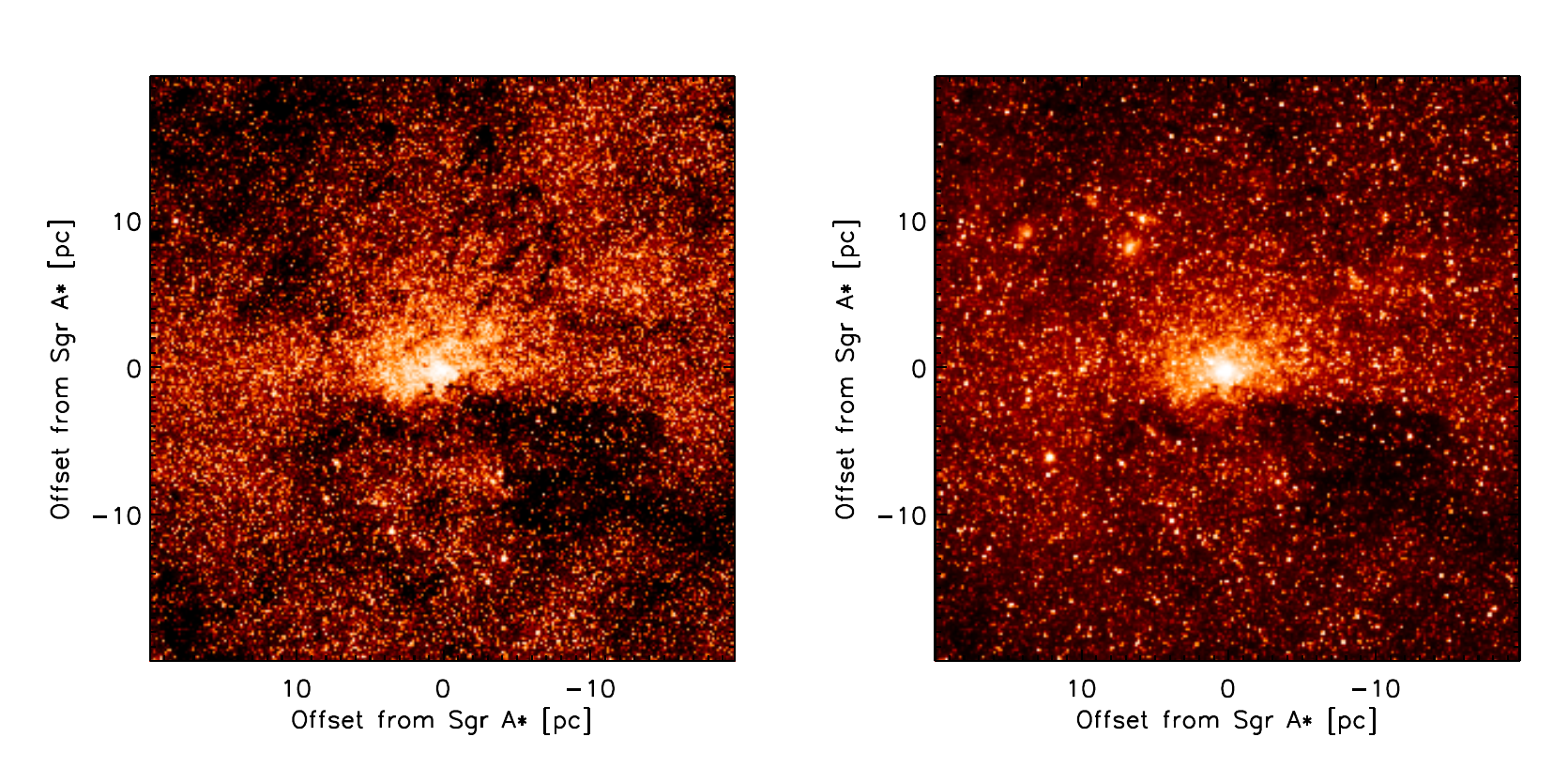}
\caption{\label{Fig:NSC_Ks_4.5}  Left: Nuclear cluster of the Milky Way at $2.15\,\mu$m seen with VIRCAM/VISTA. Right: The same field at $4.5\,\mu$m, seen with IRAC/Spitzer. Galactic north is up and Galactic east is to the left, so that the Galactic Plane runs horizontally across the middle of the images. Offsets are given in parsecs relative to Sgr\,A*. The colour scale is logarithmic and both images have been scaled in an identical way. }
\end{figure*}

Two basic scenarios are considered for the formation of NSCs, that is, inspiral and mergers of massive star clusters, such as globular clusters, and in situ formation \citep[e.g.,][]{Boker:2010ys,Hartmann:2011uq,Antonini:2013ys,Gnedin:2013vn}. It is probable that both mechanisms contribute to the growth of NSCs. The study by \citet{Seth:2006uq} of NSCs in edge-on spiral galaxies has shed light on this issue. They have found that most of the clusters in their sample are significantly flattened and closely aligned with the plane of their host galaxy. In addition, they have identified discs or rings superposed onto some of the NSCs. These additional components have a bluer colour than the actual NSCs, which suggests that the stars in the discs/rings have formed $<1$\,Ga ago. In a more detailed, integral field spectroscopy study of one of their targets, the NSC in NGC\,4244, they confirm the existence of an older, flattened spheroidal component and a younger, disc-like component. In addition, they find that the entire NSC rotates parallel to the rotation of its host galaxy \citep{Seth:2008kx}. These findings lead them to strongly favour a scenario where most of the NSC mass is formed through the infall of gas from the galaxy disc, followed by in situ star formation or by the infall of young star clusters, which are formed near the nuclear cluster, along the galaxy plane.

The biggest obstacle for studying NSCs, and galactic nuclei in general, is their great distance, which limits us to the study of the integrated light, averaged on scales of several parsecs to tens of parsecs, that is dominated by the brightest stars. Even with the next generation 30-40\,m class telescopes this situation will remain fundamentally unchanged. The centre of the Milky Way is, however, located at only $\sim$8 kpc from Earth, about a hundred times closer than the Andromeda galaxy, and five hundred times closer than the next active galactic nucleus. The Galactic centre (GC) contains the nearest NSC and MBH and is the only galaxy nucleus in which we can actually resolve the stellar population observationally and examine its properties and dynamics on scales of milli-parsecs. It is thus a crucial laboratory for studying galactic nuclei \citep{Ghez:2009kx,Genzel:2010fk,Schodel:2011ab}. In this work we assume a GC distance of $8.0$\,kpc \citep[][]{Malkin:2012uq}.

An important aspect in the study of the GC is that it presents one of the best cases for the existence of a MBH. The mass of the Galaxy's MBH,  Sagittarius\,A* (Sgr\,A*), has been measured with high accuracy through the observation of the orbits of individual stars \citep[e.g.,][]{Ghez:2008fk,Gillessen:2009qe}. The $\sim4\times10^{6}$\,M$_{\odot}$ MBH Sgr\,A* is surrounded by the Milky Way's NSC (MWNSC), the  mass of which has been estimated to $(3\pm1.5)\times10^{7}$\,M$_\odot$ \citep{Launhardt:2002nx} and its half light radius to 3-5\,pc \citep[][]{Graham:2009lh,Schodel:2011ab}. The MWNSC has been found to rotate parallel to Galactic rotation \citep{Trippe:2008it,Schodel:2009zr} and to contain multiple stellar populations \citep[e.g.,][]{Krabbe:1995fk,Paumard:2006xd,Pfuhl:2011uq}. Of particular interest is that at least 50\% of the stars that formed in the most recent star formation event \citep[between $2-6$\,Ma ago, ][]{Lu:2013fk} appear to have formed in situ in a disc around Sgr\,A* \citep{Levin:2003kx,Paumard:2006xd,Lu:2009bl}. We note that this disc, with a radius $\lesssim0.5$\,pc, is, however, much smaller than the discs or rings of young stars found in extragalactic systems \citep{Seth:2006uq,Seth:2008kx}.  This may be a selection effect from the lack of linear resolution in other galaxies, where it would be extremely hard to resolve stellar populations on scales below 0.5\,pc. On the other hand, the discs or rings observed by  \citet{Seth:2006uq,Seth:2008kx} may be more like the nuclear stellar disc (NSD) of $\sim$200\,pc radius in which the MWNSC is embedded \citep{Launhardt:2002nx}.

Our current knowledge of the MWNSC shows that it is probably a close cousin to its extragalactic counterparts and can thus serve as a benchmark for these far-away and therefore observationally unresolved systems. There are, however, significant uncertainties in our knowledge of the intrinsic properties of the MWNSC because the strong and spatially highly variable extinction across the GC region \citep[$A_{K}\approx2-5$, e.g.,][]{Scoville:2003la,Schodel:2010fk}  subjects our observations, even at near-infrared (NIR) wavelengths, to potentially significant bias. 

Although the apparent elongation of the MWNSC along the Galactic plane has already been pointed out by \citet{Becklin:1968nx}, up to now almost all observational and theoretical work has implicitly assumed a spherical shape of the cluster. This assumption was influenced in part by early work on extragalactic NSCs, which suggested that they were spherical. More recent work, however, shows that many NSCs may indeed by flattened and aligned with the disc of their host galaxies \citep[e.g.,][]{Seth:2008rr,Seth:2010fk}. Azimuthal averaging can obviously affect the estimated density law and half-light radius. In addition, most existing work has not taken the contribution from the NSD into which the MWNSC is embedded into account, which may also have biased some of the measured quantities \citep[see discussion in][]{Graham:2009lh}. The size, shape, and total mass of the MWNSC are fundamental quantities that must be accurately known  if we want to study the formation history of the GC and the interaction between the central BH and the surrounding stellar cluster. A flattened NSC, for example, would suggest formation from material that fell in predominantly along the Galactic plane. The question of a spherical or axisymmetric shape of the MWNSC can also affect our understanding of the interaction between the MBH and the surrounding stars. For example, cusp growth has so far almost exclusively been studied in spherical, isotropic systems. Intriguingly, the stellar distribution within $0.5$\,pc of Sgr\,A* is far flatter than what has been predicted by theoretical work \citep[see, e.g.,][]{Buchholz:2009fk,Do:2009tg,Bartko:2010fk}. Can this be related to erroneous assumptions about the overall properties of the MWNSC? Finally, the rate of events such as EMRIs or stellar disruptions will also depend on the overall size and shape of the MWNSC.

Fortunately, interstellar extinction is a strongly decreasing function of wavelength. Towards the GC it reaches minimum levels at mid-infrared (MIR) wavelengths of $3-5\,\mu$m \citep{Nishiyama:2009oj,Fritz:2011fk}. To illustrate this point, we show a comparison between an image of the MWNSC at $2.15\,\mu$m and at $4.5\,\mu$m in Fig.\,\ref{Fig:NSC_Ks_4.5}. It can be easily seen that the interstellar clouds near the GC, in particular towards (Galactic) south of the NSC, are almost opaque in the NIR, while they become partially transparent in the MIR.  The aim of our paper is therefore to use MIR images from the Spitzer Space Telescope from the IRAC survey of the GC \citep{Stolovy:2006fk,Arendt:2008fk,Ramirez:2008fk} to infer the intrinsic large-scale structure of the MWNSC and address the following questions: Is it spherically symmetric or flattened? In the latter case, is it aligned with the Galactic plane?  What is its size and overall luminosity?

\begin{figure}[!tb]
\includegraphics[width=\columnwidth,angle=0]{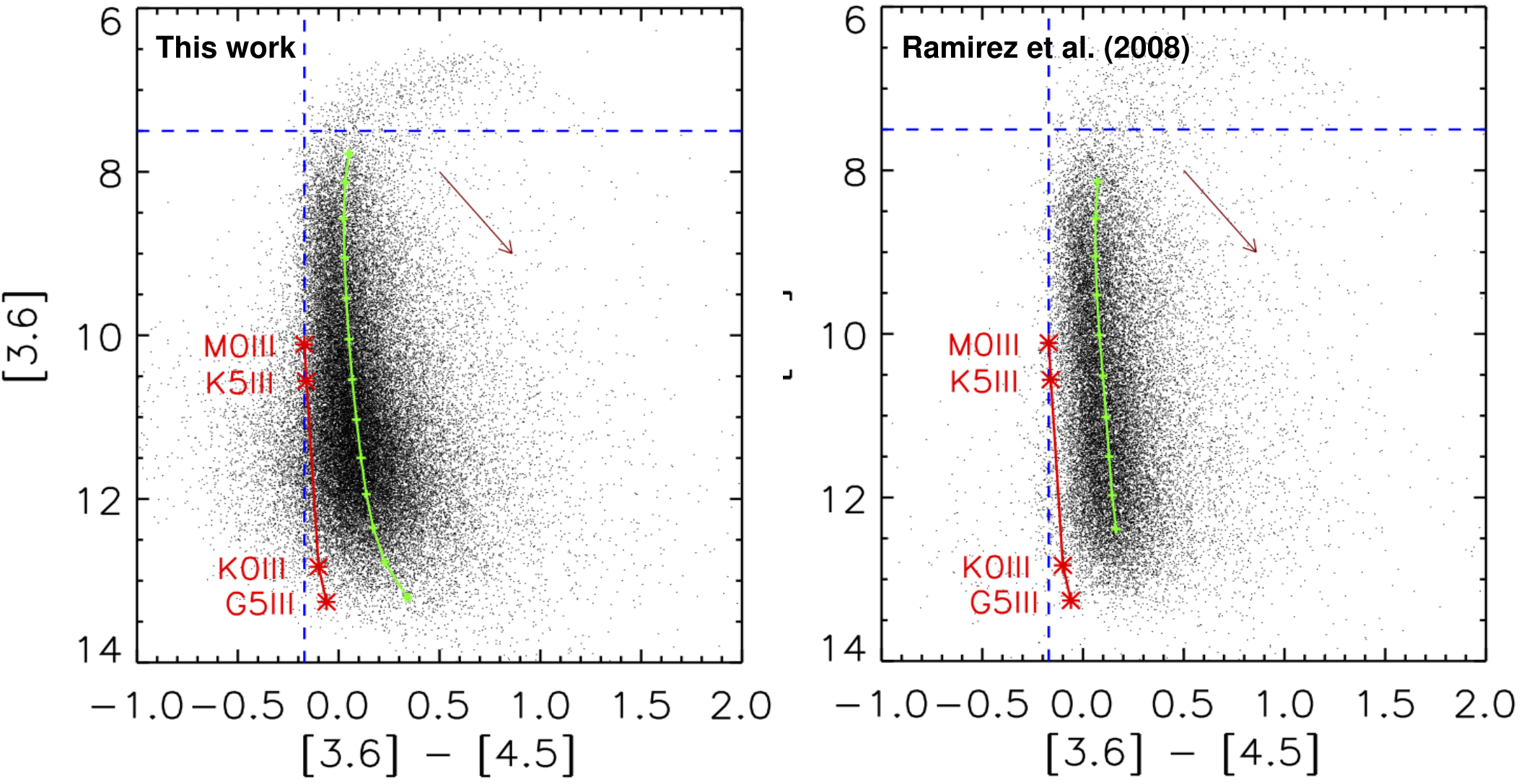}
\caption{\label{Fig:CMD}  CMD resulting from our work (left) compared to the results of \citet{Ramirez:2008fk}. The green line connects the mean colours in given magnitude intervals and the red line shows the unreddened giant sequence. The red arrow indicates the reddening law used in this work. The blue dashed lines indicate magnitude and colour cuts applied to the detected stars.}
\end{figure}


\section{Data and analysis}

Fully reduced and photometrically calibrated images of the GC region have been provided by the Spitzer/IRAC survey of the GC region by \citet{Stolovy:2006fk}.  As described in \citet{Ramirez:2008fk} and \citet{Arendt:2008fk}, regions that were heavily saturated in the main survey (with $1.2$\,s exposures) were substituted with unsaturated data taken in subarray mode (with $0.02$\,s exposures).  These regions included the central parsecs, the Quintuplet cluster, and a few other areas in the survey.  We note that the archival mosaics and GLIMPSE team mosaics do not include the subarray data. 

We chose to use the images from Channels\,1 ($3.6\,\mu$m) and 2 ($4.5\,\mu$m), where the emission is still dominated by stellar light, not by warm dust or emission from polycyclic aromatic hydrocarbons (PAHs). In addition, we used the Channel\,4 ($8.0\,\mu$m) image as a proxy to correct the shorter wavelength images  for contaminating non-stellar emission that is mostly due to PAHs. We exploit the relative uniformity of the intrinsic stellar colour measured by Channels\,1 and 2 ($[3.6-4.5]$, see below) to tackle the problem of the high and spatially variable interstellar extinction.

\subsection{Point source photometry \label{sec:photometry}}

As a first step in our analysis, we performed point-source photometry and subtraction on the IRAC maps with the point spread function (PSF) fitting program package {\it StarFinder} \citep{Diolaiti:2000qo}. The motivation was threefold: First, we want to compare the use of point-source photometry vs. the direct use of the flux per pixel in rebinned images. Second, we want to create point-source subtracted images in order to investigate and correct for the diffuse emission from PAHs in the GC. Third, we want to to assess the overall influence of the brightest stars, the only ones that can be resolved in the Spitzer images, on our results. Owing to the large field-of-view (FOV) of the original images, running {\it StarFinder} on them is computationally extremely expensive. We therefore chose to limit this analysis to a field of $2048\times2048$\,pixels (corresponding to about $80\times80$\,pc at a distance of 8\,kpc) centred on Sgr\,A*.  Since the Spitzer images were slightly under-sampled with respect to their angular resolution, they were smoothed to a point-source full width at half maximum (FWHM) of $\sim$$2.5$ pixels before feeding them into {\it StarFinder}, which relies on adequately sampled data. Smoothing alters the noise characteristics of images, but this effect is not significant for this work. This is because we used uniform weighting, assuming a constant, Gaussian noise across the image, estimated with StarFinder routines. We also note that the photon numbers in the MIR are high, so that photon noise from the source can probably be neglected. In addition, we compared the results of StarFinder runs on images smoothed with Gaussians of different FWHM and did not find any deviations on the photometry that would be of concern for the relatively low photometric precision we required. The PSF was determined by using about 20 bright, unsaturated, and relatively isolated stars distributed across the field. The PSF was iteratively improved by running {\it StarFinder} with a point-source correlation threshold of $0.9$ and a $5\,\sigma$ detection threshold and then repeating the PSF extraction, explicitly taking secondary sources into account.

The Channel\,1 and 2 images are dominated by stellar crowding for all but the brightest stars and, in addition, contain diffuse emission from the ISM that is structured down to the resolution limit. This makes their analysis challenging because, on the one hand,  the detection of spurious sources, which may be related to small-scale ISM features, must be avoided and, on the other hand, sources as faint and as confused as possible should be detected. This makes it necessary to optimize both the detection process for point sources as well as the modelling of the diffuse emission. Therefore, we performed several test runs in order to find the optimal parameters for  {\it StarFinder}.   {\it StarFinder} improves its estimate of the diffuse emission in an iterative way, but keeps the so-called {\it back\_box} parameter, which determines the angular scale of the smallest modelled diffuse features, constant during the iterations. The results of our test runs were compared to the input images by eye. We found that we could improve the performance of {\it StarFinder} on the Spitzer images significantly with the following approach. First, we ran it with  $back\_box=12$\,pixels and a high minimum correlation threshold for the acceptance of point sources ($ min\_correlation=0.9$). Then we used the map of the diffuse emission estimated in this first run as input for the second run, in which we chose $\it back\_box=2$\,pixels. This resulted in a reliable modelling of the diffuse emission so that we could relax the correlation threshold for point sources ($min\_correlation=0.7$) and apply deblending of close stars. We note that the diffuse emission is mostly dominated by faint and unresolved stars, but also contains features due to PAH emission (see below).

A list of sources common to the images from Channels 1 and 2 was subsequently created, imposing the condition that the positions of the detected stars had to coincide within one pixel (corresponding to $1"$). We did not use the point source catalogue for the IRAC/Spitzer GC survey by \citet{Ramirez:2008fk} because we could identify twice as many sources in the crowded central region. In Fig.\,\ref{Fig:CMD} we show a comparison of the colour-magnitude-diagrams (CMDs) for the central region as resulting from our analysis and the one of \citet{Ramirez:2008fk}. The two CMDs are very similar, but the CMD from our analysis with {\it StarFinder} shows a weak trend towards redder colours for faint stars, as well as a larger scatter. This indicates that the point-source photometry from the catalogue of \citet{Ramirez:2008fk} is probably more accurate. However, for studying the diffuse emission and for examining the densest parts of the MWNSC, it is essential to maximise point-source detection. For this reason we use our  {\it StarFinder}-based photometry in the following

\begin{figure}[!htb]
\includegraphics[width=\columnwidth,angle=0]{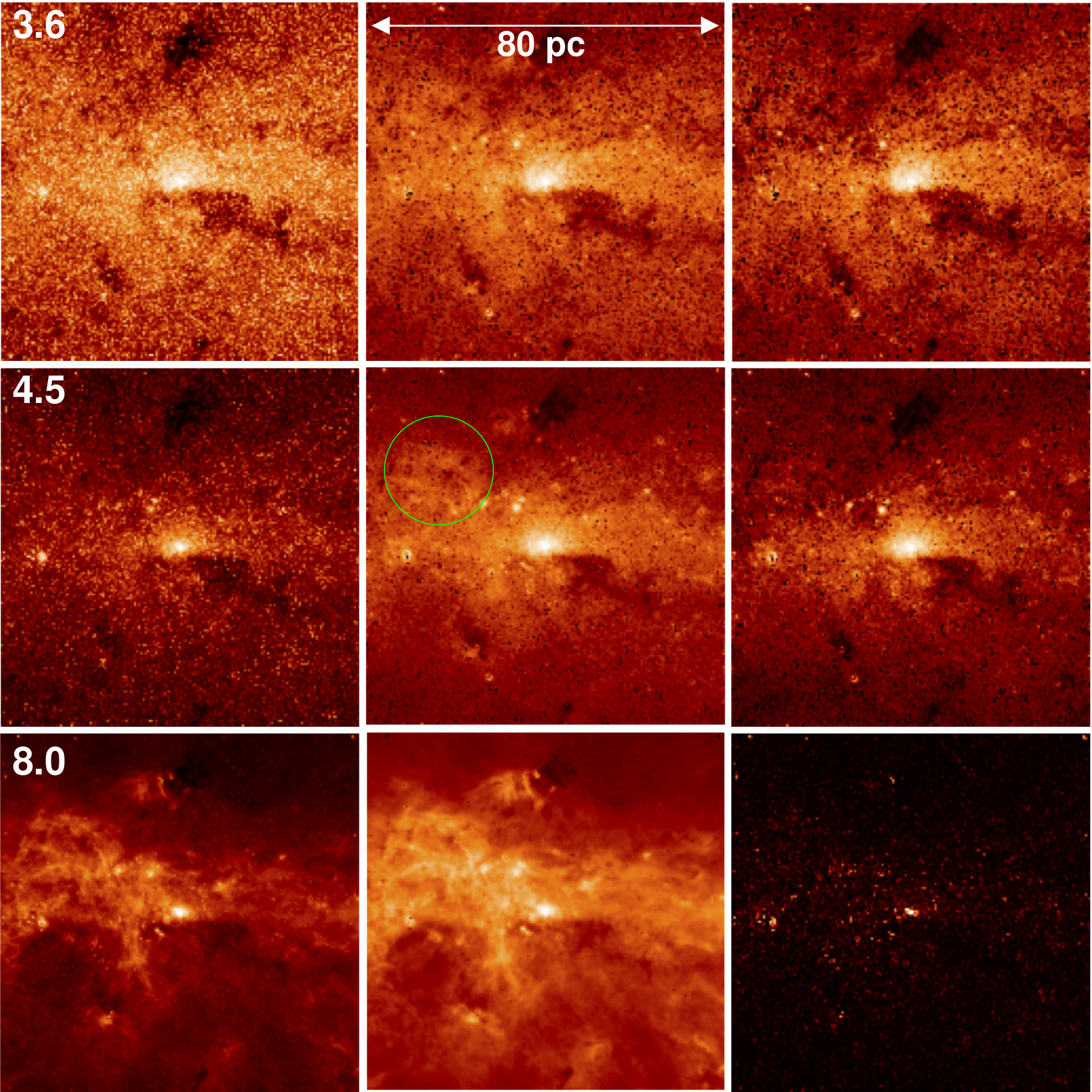}
\caption{\label{Fig:maps}  Left column: $3.6$, $4.5$, and $8.0\,\mu$m images of the central  $80\times80$\,pc$^{2}$ of the GC. The middle column shows the diffuse background resulting after point-source subtraction from each image. The right column shows, from top to bottom, the diffuse $3.6\,\mu$m emission (top middle) minus $0.03$ times the $8.0\,\mu$m image (bottom left), the  diffuse $4.5\,\mu$m emission (centre) minus $0.04$ times the $8.0\,\mu$m image (bottom left), the $8.0\,\mu$m image (bottom left) minus diffuse emission at $8.0\,\mu$m (bottom middle). The green circle in the middle panel indicates a region with strong PAH emission in the $4.5\,\mu$m image.}
\end{figure}

\subsection{Diffuse emission \label{sec:diffuse}}

As a next step, we created point-source subtracted maps at $3.6\,\mu$m, $4.5\,\mu$m, and $8.0\,\mu$m to explore the role of the diffuse emission. We rebinned the point-source subtracted images to a pixel scale of $5"$ per pixel and applied sigma-filtering with a $5\,\sigma$ threshold in a seven-pixel wide box. This procedure served to suppress artefacts or very small-scale features, such as extremely bright or saturated  individual stars or  residual features from the point-source subtraction. The original rebinned images and the corresponding maps after point-source subtraction are shown in the left and middle columns of Fig.\,\ref{Fig:maps}. The morphology of the  diffuse emission changes clearly from the shortest wavelength, where the overall emission is dominated by the stars, to the longest wavelength, where the total emission is completely dominated by the diffuse component. At $8.0\,\mu$m the diffuse emission, which is thought to be mainly due to PAHs \citep{Stolovy:2006fk,Arendt:2008fk},  amounts to $>90\%$ of the total emission and can thus be used as a reasonable proxy for the total non-stellar diffuse emission, without the need to account for point sources.

\citet{Stolovy:2006fk} and \citet{Arendt:2008fk} point out the high uniformity of the ratio of the diffuse emission at the GC between $5.8\,\mu$m and $8.0\,\mu$m. \citet{Arendt:2008fk} derive a median ratio of $I_{ISM}(5.8\,\mu\mathrm{m})/I_{ISM}(8.0\,\mu\mathrm{m}) =0.08\pm0.005$, but do not examine this ratio for the shorter wavelengths where stellar emission dominates the diffuse light. To assess the contribution of the PAH emission to the diffuse background at the shorter wavelengths, we compared the flux ratios of the diffuse emission at $8.0\,\mu$m to the ones at $3.6$ and $4.5\,\mu$m, respectively. This comparison was done in regions $>20$\,pc north and south of the Galactic plane because at lower latitudes, where the stellar density is higher, the diffuse emission is strongly dominated by unresolved stellar emission at the shorter wavelengths. We focused on regions where diffuse emission with similar morphology can be seen in all maps (see, e.g., map of the diffuse $4.5\,\mu$m emission in the central panel of Fig.\,\ref{Fig:maps}). In this way we estimated the contribution from PAHs to the total diffuse emission at $3.6$ and $4.5\,\mu$m as $0.03$ and $0.04$ times the diffuse flux at $8.0\,\mu$m. The estimated $1\,\sigma$ uncertainty of both factors is $0.005$.

We note that the uniformity of the flux ratio is difficult to test at shorter wavelengths and near the Galactic plane, where the stellar component dominates the diffuse emission, but the assumption of uniformity seemed to hold in the areas examined by us. 

\begin{figure}[!htb]
\includegraphics[width=\columnwidth,angle=0]{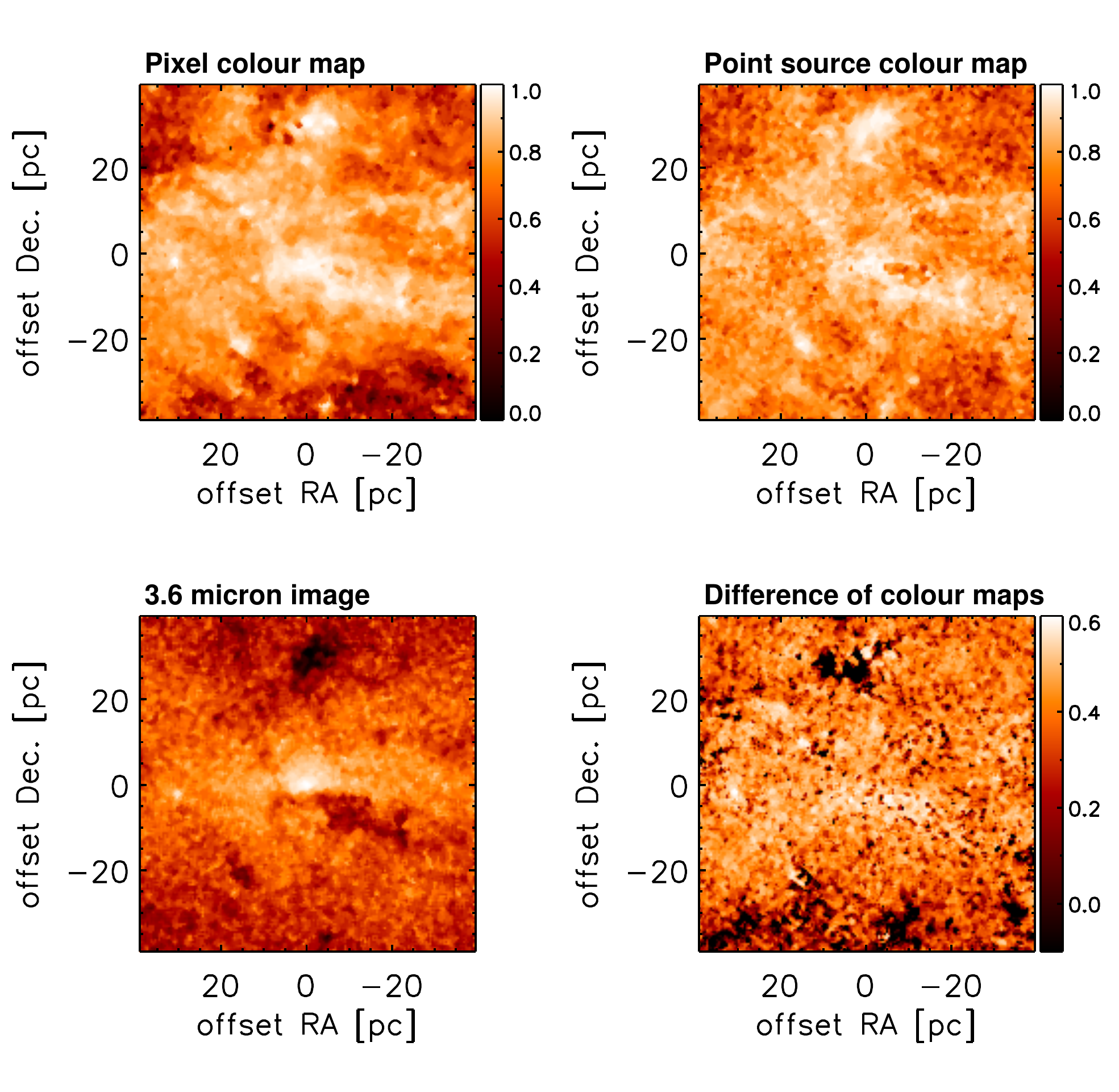}
\caption{\label{Fig:colorcheck} Upper left: Colour map obtained directly from the calibrated $4.5$ and $3.6\,\mu$m images. Upper right: Colour map obtained from the colours of stars identified via PSF fitting in the $4.5$ and $3.6\,\mu$m images. An average intrinsic colour of  $[3.6-4.5]=-0.15$\,mag (see Section\,\ref{sec:extinction}) was subtracted from both maps. Lower left: Median-smoothed $3.6\,\mu$m image. Lower right: Difference of left upper map minus right upper map. }
\end{figure}

\subsection{Colour maps}
 
Subsequently, a map of the mean $[3.6-4.5]$ colour was created. Since several stars per pixel are necessary for a reliable colour estimate, we chose a  $12"/$pixel scale for this map. Probable foreground stars, with $[3.6-4.5]<-0.17$ (see Section\,\ref{sec:extinction}), and stars above the saturation/linearity limit, with  $[3.6]<7.5$,  were excluded  (see Fig.\,\ref{Fig:CMD}).

The colour map created from the PSF photometry on stars, subsequently called {\it point source colour map}, was then compared with a colour map created directly from the images, after rebinning them  to $12"$ per pixel, subsequently called a {\it pixel colour map}. Iterative sigma filtering with a filter box of seven pixels width and a threshold of $5\,\sigma$ was applied to the rebinned maps to correct pixels that were dominated by individual, saturated, or extremely bright stars (which are frequently foreground stars) or by very compact diffuse emission. The appropriately scaled (see above) $8.0\,\mu$m image (the full image, not the point-source-subtracted one) was then subtracted from the $3.6\,\mu$m and $4.5\,\mu$m images to remove the contribution from PAHs. Fluxes were converted into magnitudes by using the zero magnitude definitions from the Spitzer IRAC Instrument Handbook.

\begin{figure}[!htb]
\includegraphics[width=\columnwidth,angle=0]{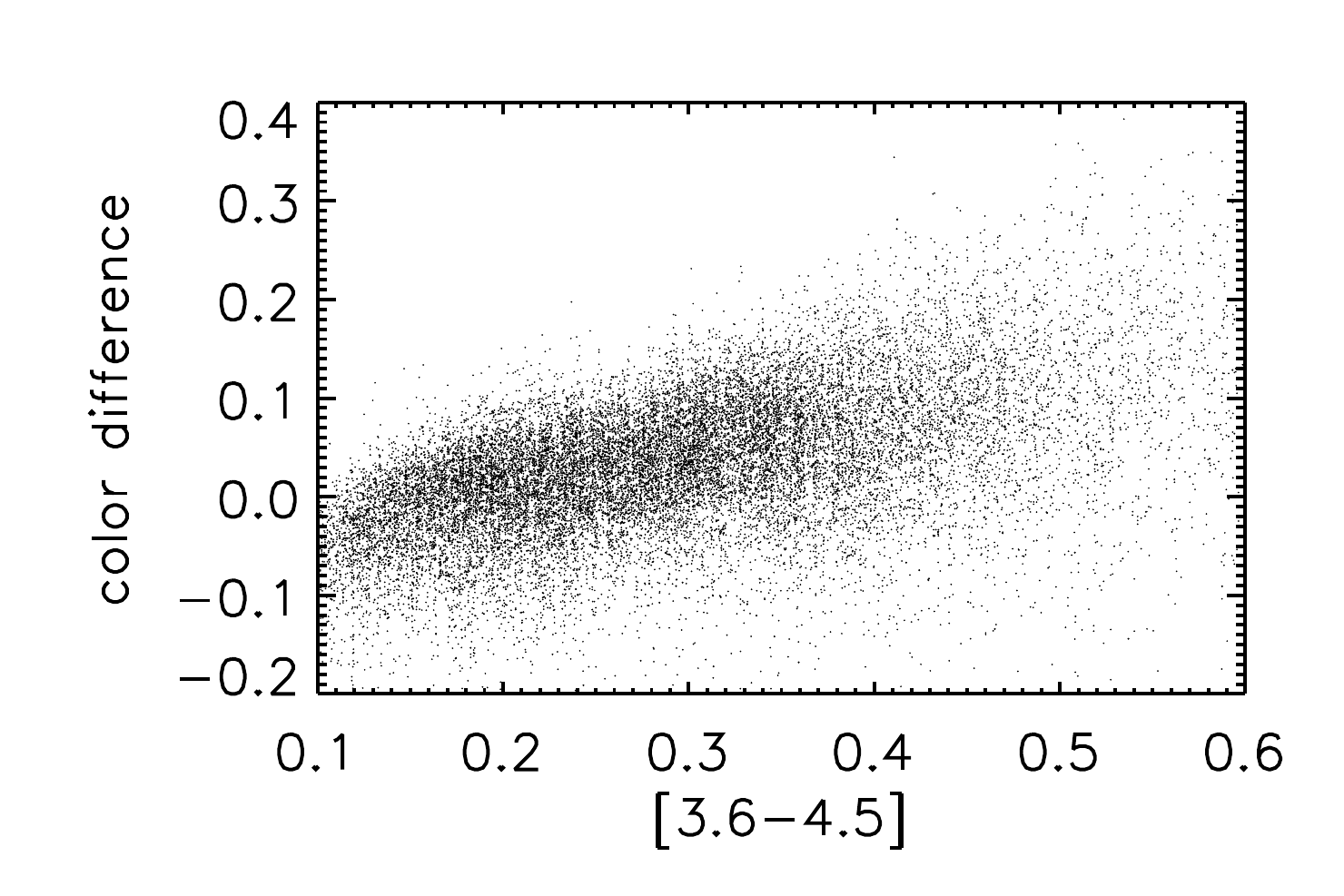}
\caption{\label{Fig:colortrends} Pixel-by-pixel colour difference between the pixel colour map and the point source colour map plotted vs. the colour in the pixel colour map.}
\end{figure}

Subsequently, we compared the point source colour map with the pixel colour map (Figs.\,\ref{Fig:colorcheck} and \ref{Fig:colortrends}). Both maps look very similar and trace the obvious dark clouds and extinction variations in the $3.6\,\mu$m image well. The point source colour map displays more small-scale variation, which is almost certainly due to noise introduced by the small number of stars per pixel (between 0 and 6; pixels containing fewer than two stars were interpolated like bad pixels). The colour difference map (bottom right panel) shows that, close to the Galactic plane and in regions with redder colours, the pixel colour map shows significantly redder colours. This is not surprising because of the effects that crowding and high extinction have on the detection of stars. Both will favour the detection of stars that are, on average, closer to the observer and thus brighter and less extinguished. The pixel colour map, on the other hand, contains the flux from all stars within a pixel, which includes the diffuse emission that is dominated by unresolved or faint (and thus generally more reddened) stars. This point is examined further in Fig.\,\ref{Fig:colortrends}, where we see that the  colour difference between the two maps is close to zero for small reddening, but increases clearly towards redder colours. 

From this comparison we conclude that using pixel colours, i.e. colours directly estimated from rebinned, flux-calibrated images that were corrected for PAH emission, is less prone to bias and more accurate in crowded and highly extinguished regions. In particular when we want to measure the total emission from the GC region, we have to be careful to apply an extinction correction that is derived from the total light, and not just from the detectable stars. We note that the comparison of the colours from PSF photometry with the pixel colours shows that both methods agree very well for low reddening, with a mean colour offset (pixel map minus star map) of just $[3.6-4.5]_{\rm offset}=0.01$ for pixels with $0.1<[3.6-4.5]<0.3$.

\subsection{Extinction map \label{sec:extinction}}

\begin{figure*}[!ht]
\centering
\includegraphics[width=.8\textwidth,angle=0]{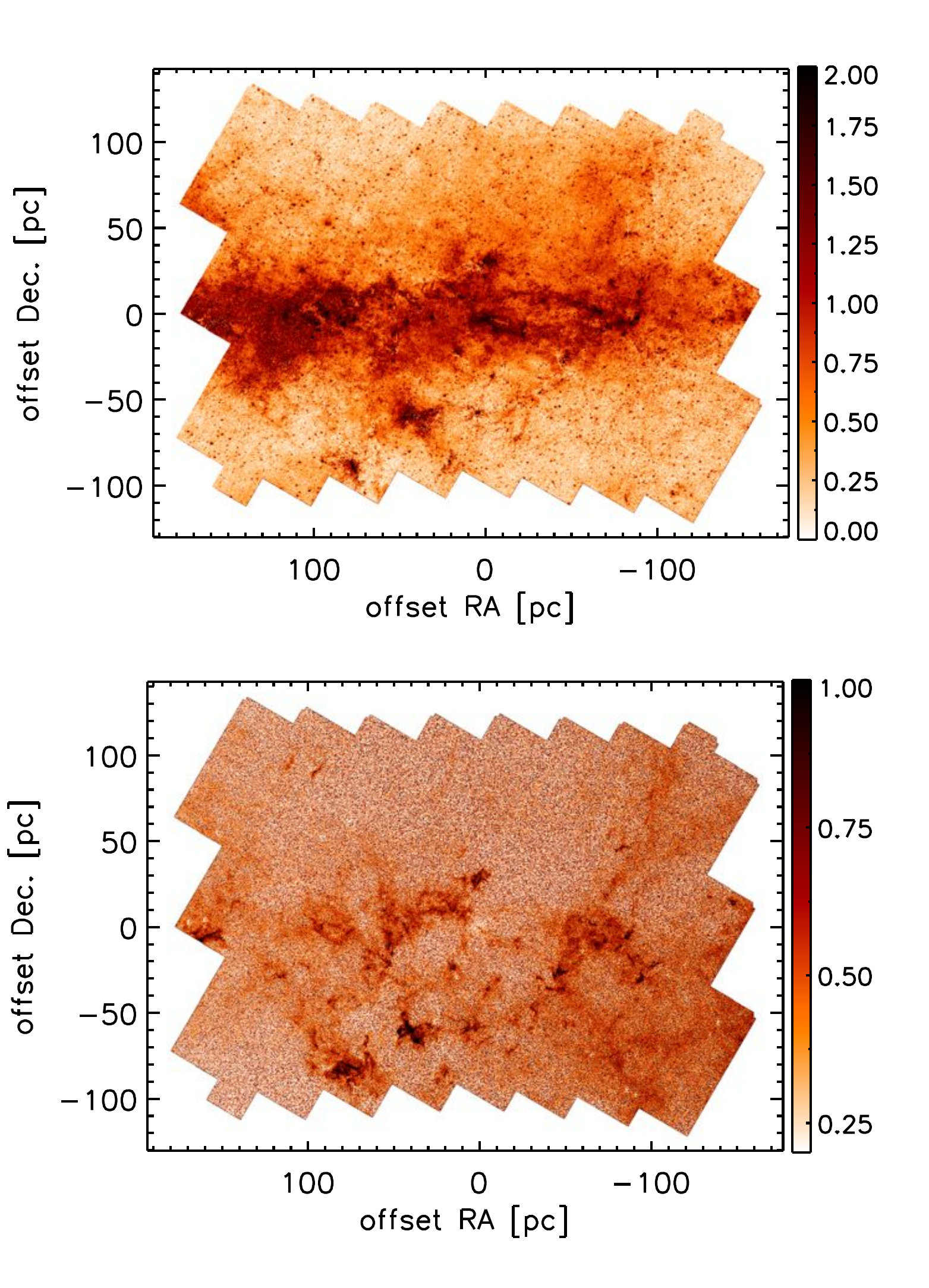}
\caption{\label{Fig:ext} Upper panel:  $4.5\,\mu$m extinction map. Lower panel: Corresponding uncertainty map, including statistical and systematic uncertainties. The unit of the colour bar is in magnitudes. These maps are made available in electronic form at the CDS.}
\end{figure*}
 
The tests described in the section above show that we can obtain reliable estimates of the colour of the stellar emission toward the GC by using directly the calibrated IRAC/Spitzer maps. Subsequently, we proceeded as follows. First, the maps were rebinned to a pixel scale of $5"$ per pixel, i.e. to a higher resolution than what is possible with the point source maps, but still low enough to allow us to get an accurate estimate of the mean flux and its uncertainty for each pixel. After rebinning, small-scale artefacts, such as saturated stars or individual bright stars, could be  removed effectively by sigma filtering with a $5\,\sigma$ threshold in a seven-pixel-wide box. The flux uncertainty of each pixel in the rebinned maps was estimated from the uncertainty of the mean flux of all corresponding pixels from the original images. The contribution from PAHs to the diffuse flux was corrected for by subtracting the $8.0\,\mu$m map scaled with factors $0.03$, for the $3.6\,\mu$m image, and $0.04$, for the $4.5\,\mu$m image (see Section\,\ref{sec:diffuse}). A colour map was finally created by subtraction of the calibrated images.

The intrinsic stellar colours,  $[3.6-4.5]_{0}$,  were estimated by combining the stellar atmosphere models of \citet{Kurucz:1993fk} with the corresponding IRAC filter transmission curves. We found that the stellar colours were small in this wavelength regime, varying between $-0.06$ and $-0.17$ for giants. The dominating population of K and M giants have colours around $-0.15$. This value agrees closely with the colours computed by \citet{Majewski:2011uq} for late-type stars (see their Fig.\,3) and was adopted as the intrinsic colour of the mean stellar emission, which was then subtracted from the colour map.

To convert colours into extinction, we initially used the extinction law derived by \citet{Nishiyama:2009oj}. They found $A_{[3.6]}/A_{[4.5]}=0.50/0.39=1.28/1$, corresponding to $A_{[4.5]}/E[3.6-4.5]  = 1/0.28 = 3.57$, where $E[3.6-4.5] = [3.6-4.5] - ([3.6 - 4.5])_{0}$ is the colour excess. The interstellar extinction for each pixel was thus computed as $A_{4.5\,\mu m}=E[3.6-4.5] \times 3.57$. When applying the extinction map derived in this way to the correspondingly rebinned IRAC $4.5\,\mu$m map, we found that this resulted in over-correction, i.e., extinguished areas appeared as excess emission after correction. This is not necessarily surprising because assuming a flat extinction law in the MIR is an over-simplification. A close inspection of available measurements and models for the optical-to-mid-infrared (MIR) extinction shows that  (a) a power law can only serve as a rough approximation in the spectral range considered here and  (b) if we assume a power law, its exponent is poorly constrained by the available data. In addition, the power law exponent will depend strongly on the central wavelengths and widths of the filters used because of strong features due to aliphatic hydrocarbons near $3.4\,\mu$m or  $CO_{2}$ and $CO$ ice at $4.3\,\mu$m and $4.7\,\mu$m, respectively \citep[see, e.g., Fig.8 in ][]{Fritz:2011fk}. The IRAC Channel\,1 has an effective wavelength of $3.550\,\mu$m with a bandwidth of $0.750\,\mu$m and Channel\,2 has an effective  wavelength of $4.493\,\mu$m with a bandwidth of $1.015\,\mu$m (see Spitzer IRAC Instrument Handbook). This means that the extinction law that must be adopted depends strongly on the central wavelengths and widths of the used broad-band filters. We therefore had to determine the value of  $A_{[4.5]}/E[3.6-4.5]$ again specifically for our case. This was done by checking by eye the extinction-corrected emission maps. After correction, extinguished areas should neither be brighter nor fainter than their surroundings. We thus estimated  $A_{[4.5]}/E[3.6-4.5]=1.8$ with a $1\,\sigma$ uncertainty of $0.2$  for our measurements.

The upper panel of Fig.\,\ref{Fig:ext} displays the resulting extinction map at $4.5\,\mu$m. It traces very similar features to the opacity (atomic hydrogen density) map derived by \citet{Molinari:2011fk} from Herschel far-infrared observations.  This good agreement gives us confidence in the accuracy of our extinction map. The uncertainty map corresponding to the extinction map is shown in the  lower panel  of Fig.\,\ref{Fig:ext}. The uncertainty map, $\Delta Ext$, was calculated as $\Delta Ext = \sqrt{\Delta Ext_{col}^{2} + \Delta Ext_{\alpha}^{2} + \Delta Ext_{PAH}^{2}}$ from the individual statistical and systematic uncertainties that derive from the uncertainty of the measured colour of the stellar emission ($\Delta Ext_{colour}$), the uncertainty of the extinction law ($\Delta Ext_{\alpha}$), and the uncertainty of the PAH correction ($\Delta Ext_{PAH}$). Generally, $\Delta Ext_{colour} > \Delta Ext_{alpha} > \Delta Ext_{PAH} $. To give an idea of the overall magnitude of the uncertainties,  their median values are $Median(\Delta Ext_{col}) = 0.25$, $Median(\Delta Ext_{\alpha}) = 0.01$, and $Median(\Delta Ext_{PAH}) = 0.001$. 

\subsection{MIR maps of the stellar emission}

\begin{figure}[!tb]
\includegraphics[width=\columnwidth,angle=0]{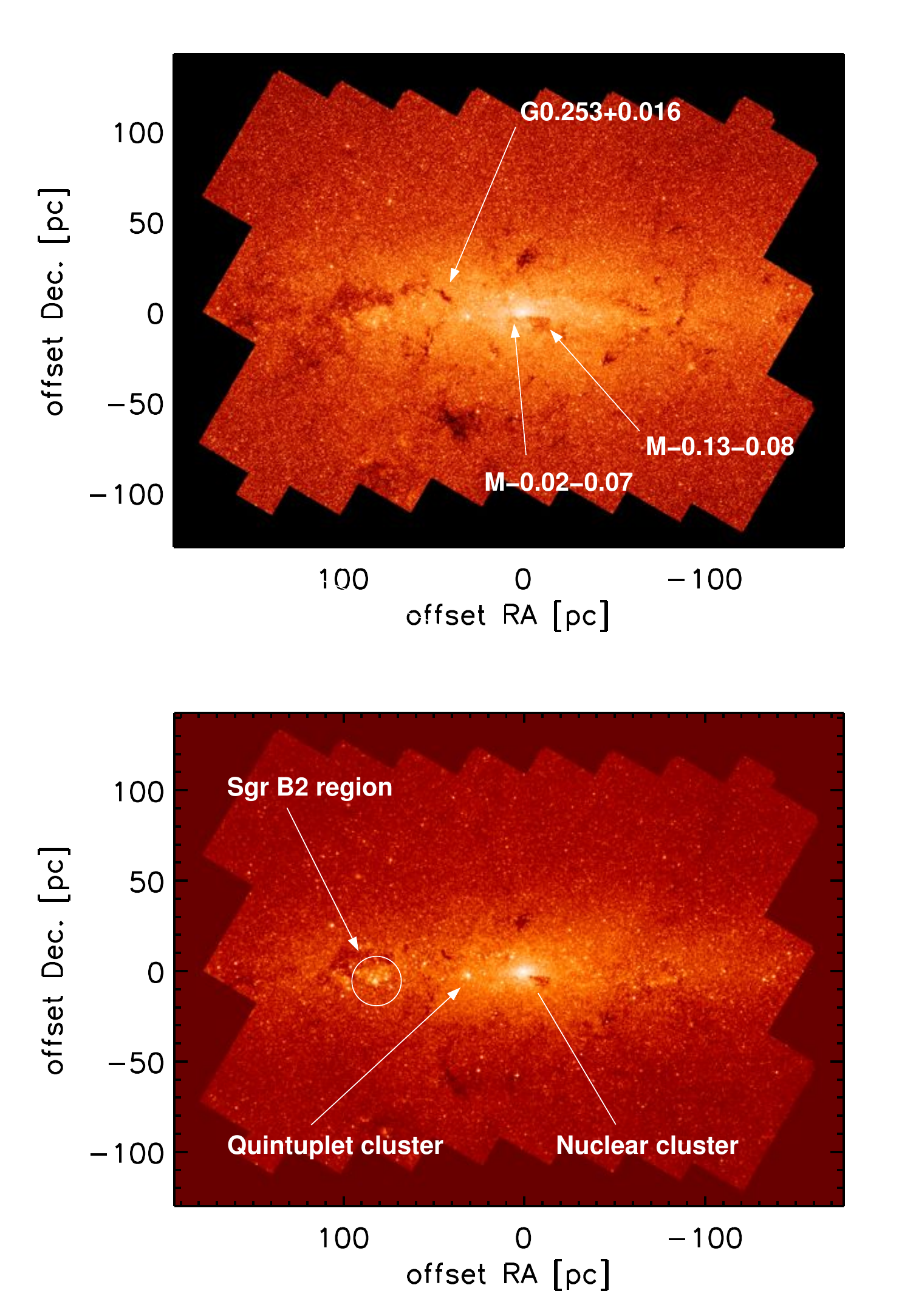}
\caption{\label{Fig:corrmaps} Upper panel: IRAC $4.5\,\mu$m image of the GC region, after subtracting of the estimated  diffuse emission from the PAHs.  Lower panel: Image as in upper panel, but corrected for extinction.}
\end{figure}

The extinction map was used to correct the PAH-subtracted $4.5\,\mu$m map, which is shown before and after extinction correction in Fig.\,\ref{Fig:corrmaps}. Several infrared dark clouds within the field show up as dark clouds even after extinction correction, i.e., our map only provides lower limits for the extinction toward these clouds. Such dark clouds are, for example, the well known M-0.13-0.08, also known as the 20 km\,s$^{-1}$ cloud \citep[see, e.g., Fig.\,1 in][]{Garcia-Marin:2011fk}, or the extremely dense molecular clump G.0253+0.016 \citep[e.g., ][]{Longmore:2012uq}. M-0.02-0.07, the so-called 50 km\,s$^{-1}$ cloud \citep[see Fig.\,1 in][]{Garcia-Marin:2011fk}, on the other hand, appears to be largely transparent to MIR light. 

The two major stellar structures that dominate the extinction-corrected IRAC $4.5\,\mu$m map are the nuclear stellar disc (NSD) and the NSC that together form the so-called nuclear bulge of the Milky Way \citep[][]{Launhardt:2002nx}. The Sgr\,B2 star-forming region and the Quintuplet cluster show prominent local compact excess emission that stands out from the surrounding NSD. The younger Arches cluster, which contains less evolved stars, does not show up as a prominent source.

\section{Structure of the nuclear star cluster \label{sec:structure}}

\begin{figure}[!htb]
\includegraphics[width=\columnwidth,angle=0]{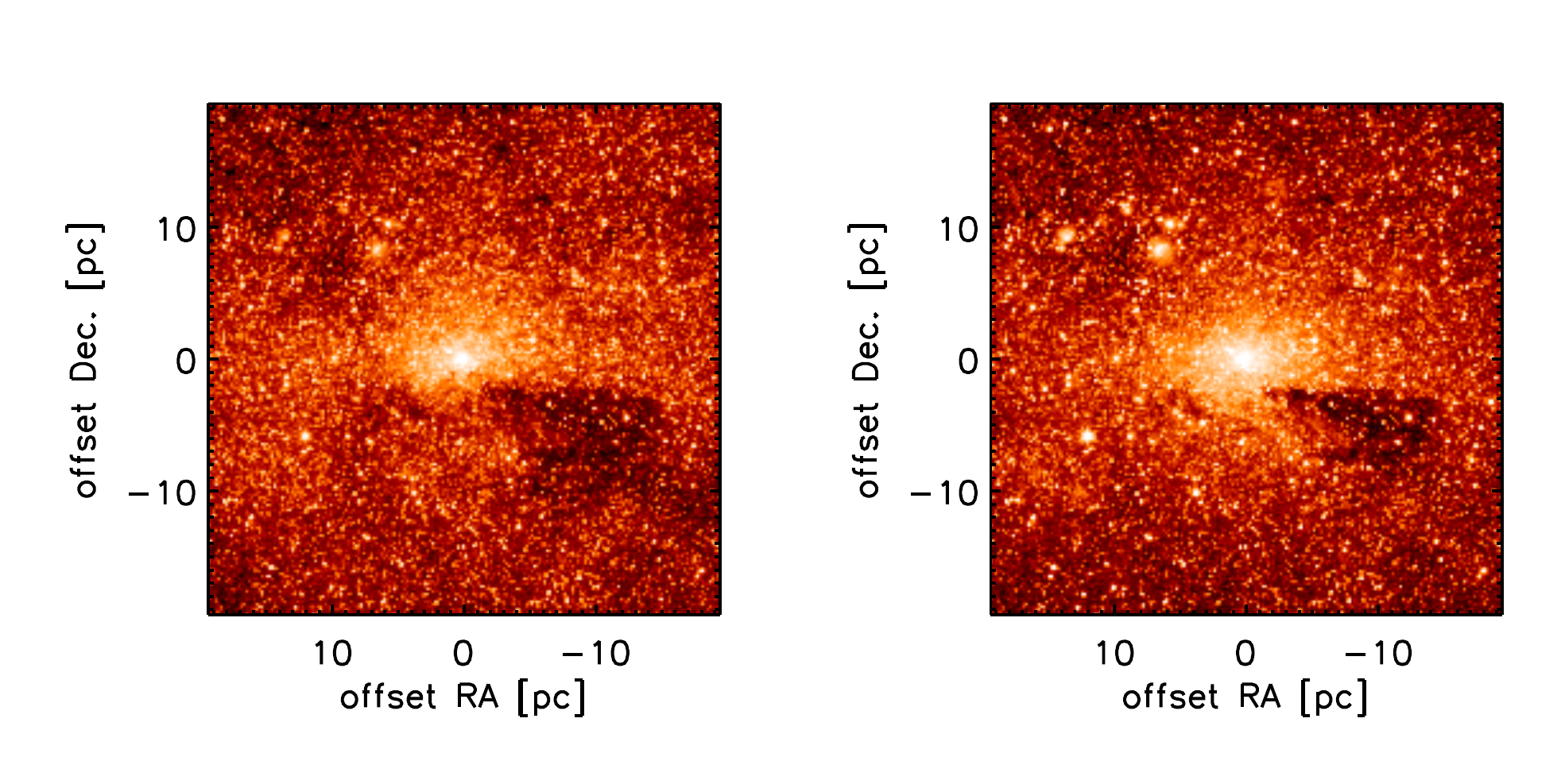}
\caption{\label{Fig:NSC}  Zoom onto the MWNSC in the IRAC $4.5\,\mu$m image. Left panel: PAH-corrected image. Right panel: PAH and extinction-corrected image. Galactic north is up and Galactic east is to the left.}
\end{figure}

A close-up of the  PAH-subtracted $4.5\,\mu$m image of the NSC is shown before and after  extinction correction in Fig.\,\ref{Fig:NSC}. The only large structures apart from the NSC that are visible in the corrected image are four HII regions about 15\,pc to the northeast of Sgr\,A* (located at the origin) and the large infrared dark cloud M-0.13-0.08 (the 20\,km\,s$^{-1}$ cloud) to the southwest. The HII regions may be dominated by a strong radiation field or a lack of PAHs and therefore deviate from the assumed homogeneous colour of the PAH emission \citep{Arendt:2008fk}.  

\begin{figure}[!htb]
\includegraphics[width=\columnwidth,angle=0]{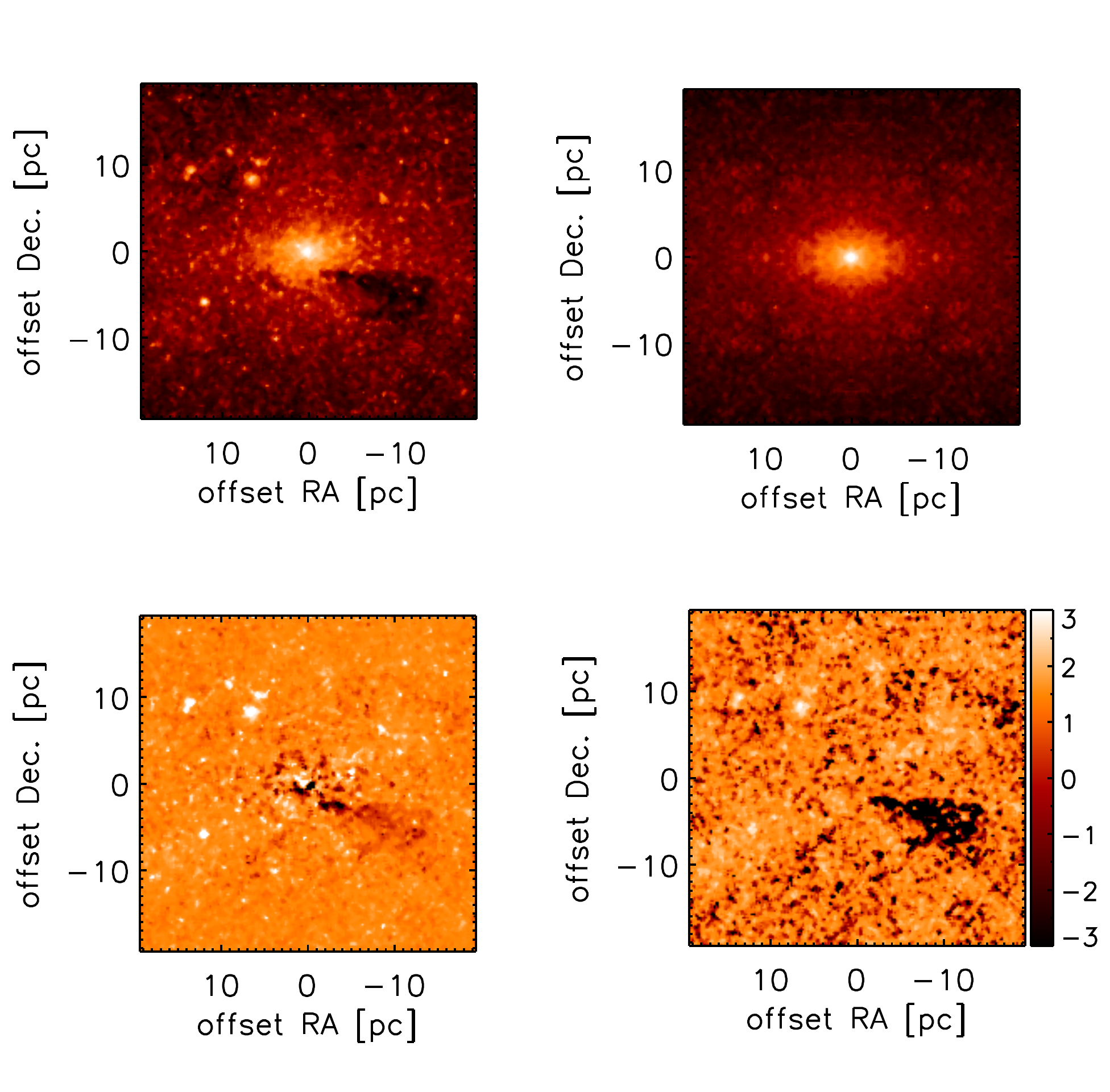}
\caption{\label{Fig:symmetry} Top  left: Zoom onto PAH- and extinction-corrected image of the MWNSC at $4.5\,\mu$m. Top right: Symmetrised image resulting from folding and median-averaging the four quadrants. Both images in the top have been median-smoothed with a 3-pixel wide box to enhance the large-scale structures. Bottom left: Difference between top left and right panels. Bottom right: Difference image divided by the uncertainty map resulting from quadratically combining the pixel uncertainties of the images in the top left and right panels. Galactic north is up and Galactic east is to the left. The colour bar in the bottom right panel is in units of standard deviations. }
\end{figure}

It is obvious that even after extinction correction the NSC still appears to be extended along the Galactic plane.   To appreciate the large-scale structure of the NSC better, a median-smoothed version (using a 3-pixel/$0.6$-pc-wide box) of the corrected image is shown in the top left panel of Fig.\,\ref{Fig:symmetry}. The top right panel shows a folded image of the NSC that was obtained by assuming symmetry of the cluster with respect to the Galactic plane and with respect to the Galactic north-south axis through Sgr\,A*. The folded image was obtained by replacing each pixel in each image quadrant with the median of the corresponding pixels in the four image quadrants. The uncertainty for each pixel was computed from the uncertainty of the mean. The pixels along the vertical and horizontal symmetry axes of the image were not averaged.  If the NSC does indeed possess the previously assumed symmetry, then we do not expect any strong residuals in the difference image between the original and the folded/symmetrised image of the NSC. The bottom left-hand panel of Fig.\,\ref{Fig:symmetry} shows such a difference map. Significant residuals can only be seen in the north-east quadrant, where the previously mentioned HII regions lie, in the south-west quadrant, where the dark cloud M-0.13-0.08 is located, and near the centre. After normalising each pixel by its uncertainty (computed from a quadratic combination of the uncertainties of the pixels in the original and in the folded map) the residuals near the centre disappear, as is shown in the bottom right-hand panel of Fig.\,\ref{Fig:symmetry}. With the exception of the strong systematic residual caused by the dark cloud M-0.13-0.08 and some excess emission near the previously mentioned HII regions in the north-east quadrant,  we do not detect any significant systematic deviations. This supports our assumptions about the symmetry of the MWNSC.

\begin{figure}[!htb]
\includegraphics[width=\columnwidth,angle=0]{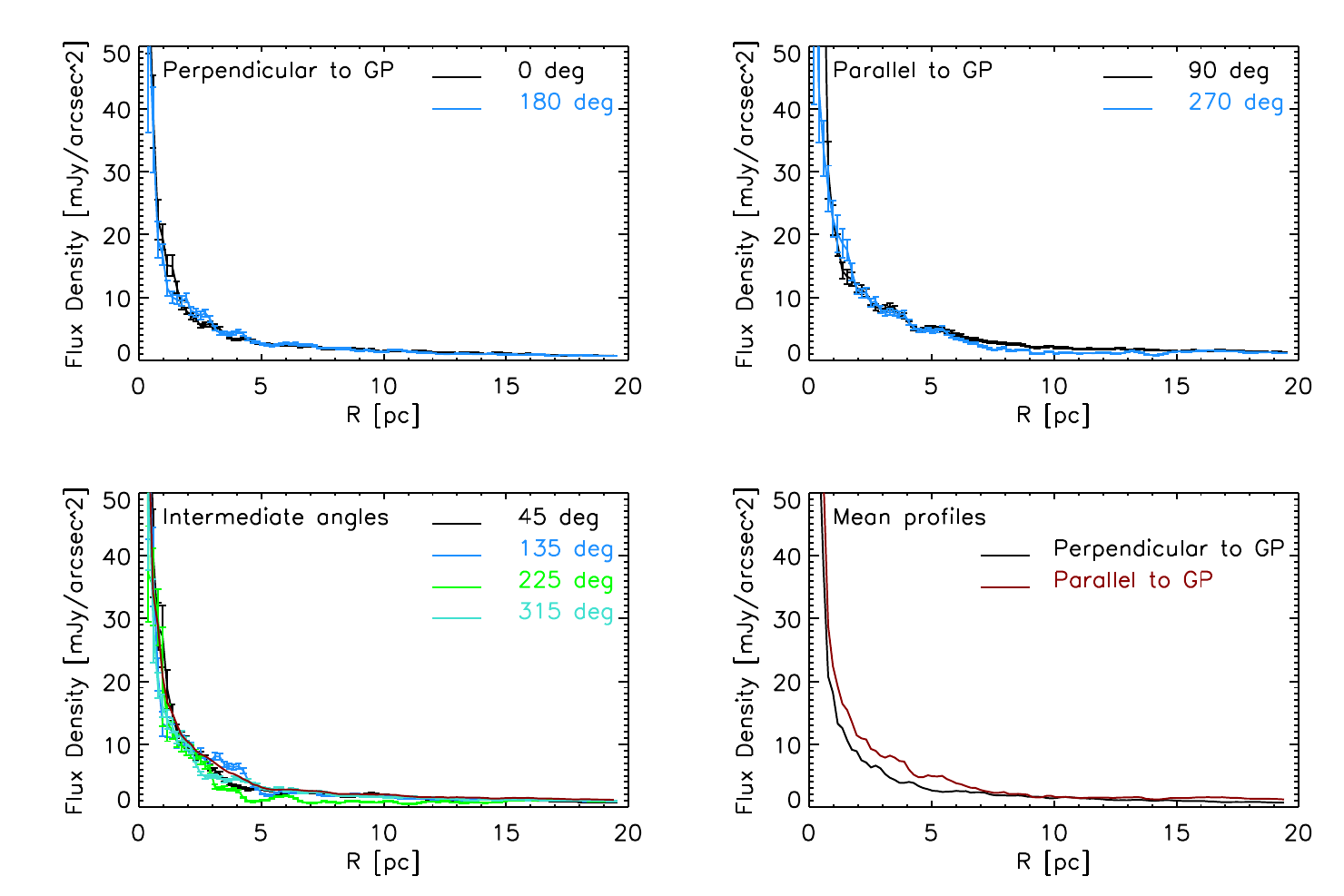}
\caption{\label{Fig:profiles} Top left: Profiles of the flux density of the NSC in 45 degree-wide wedges perpendicular to the Galactic plane (GP). Top right: Profiles parallel to the GP. Bottom left: Profiles at intermediate angles. The profile at $225^{\circ}$ is systematically lower at $5-15\,$pc because of the dark cloud southwest of the NSC.  Bottom right: Mean profiles parallel and perpendicular to the GP. Angles increase east of north, with $0^{\circ}$ corresponding to Galactic north.}
\end{figure}

\begin{figure}[!htb] \includegraphics[width=\columnwidth,angle=0]{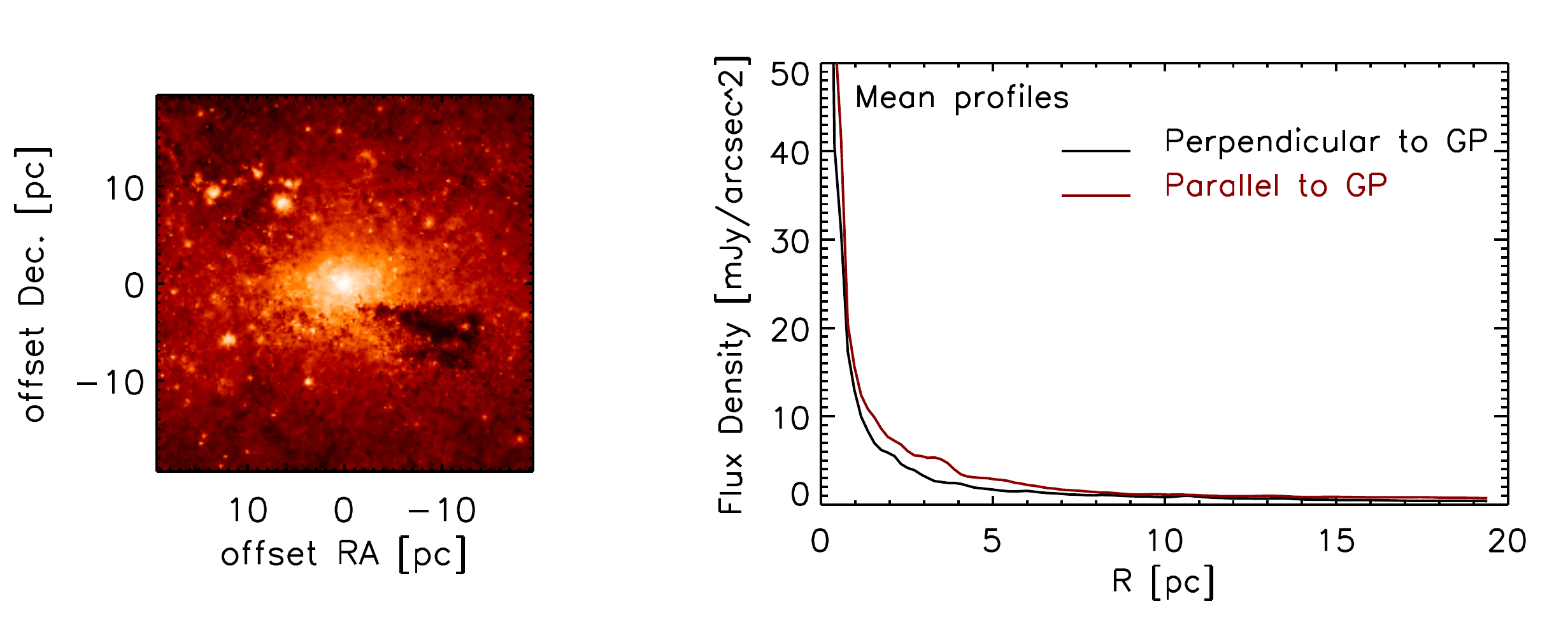} \caption{\label{Fig:diffuse} Left: Image of the diffuse $4.5\,\mu$m emission of the NSC, after subtraction of the point sources detected by {\it StarFinder}. The image was corrected for PAH emission and extinction. Right: Profile of the mean diffuse emission parallel and perpendicular to the Galactic plane.}
\end{figure}

To further investigate the symmetry of the MWNSC, we calculated flux-density profiles on the corrected $4.5\,\mu$m image in conical sections of $45^{\circ}$ width that have their apices fixed on Sgr\,A* (similar to wedges in a pie chart). The wedges are oriented at steps of $45^{\circ}$ with $0^{\circ}$ pointing northwards and increasing counter-clockwise. The angles of $90^{\circ}$ and $270^{\circ}$ correspond therefore to directions east and west within the Galactic plane and $0^{\circ}$ and $180^{\circ}$ to north and south. The profiles, shown in Fig.\,\ref{Fig:profiles} are similar  within their uncertainties for corresponding directions, but are markedly different parallel and perpendicular to the Galactic plane (bottom right panel). These results agree with the hypothesis of a point-symmetric cluster that is flattened in the Galactic north-south direction, in agreement with its rotation parallel to Galactic rotation \citep{Trippe:2008it,Schodel:2009zr}.

The IRAC/Spitzer MIR images of the GC are dominated by crowding and are incomplete at all but the brightest stellar magnitudes, in particular in the MWNSC region. Also, small numbers of extremely bright stars, for example IRS\,7 in the central parsec, may bias the measured flux distribution and thus the inferred shape of the MWNSC (although our rebinning and sigma-filtering procedure largely suppresses the influence of individual bright stars). We therefore performed an additional test on the point-source subtracted $4.5\,\mu$m image (see Section\,\ref{sec:diffuse}). The same processing (rebinning and filtering) and  correction (PAH and extinction) steps were applied as to the original $4.5\,\mu$m image. The result is shown in the left-hand panel of Fig.\,\ref{Fig:diffuse}. The corresponding profiles of the diffuse, unresolved flux density averaged along and perpendicular to the Galactic plane (GP)  are shown in the right-hand panel. The overall agreement with the appearance of the NSC as shown in Fig.\,\ref{Fig:NSC} and with the profiles displayed in Fig.\,\ref{Fig:profiles} is very good. We thus conclude that our results cannot be significantly biased by a small number of individual bright sources.

\section{Model fits}
\label{sec:models}

Since we are not interested in the NSD here, which will be examined in an upcoming paper (Kunneriath et al., in prep.), we primarily focussed on a region of about $20\times20$\,pc, centred on Sgr\,A*, and used different methods to estimate the contribution of the stellar fore- and background emission. We then fit elliptical King, Moffat, and S\'ersic functions to the image of the NSC.  With the exception of the symmetrised images, where it was not necessary, we masked the 20\,km\,s$^{-1}$ cloud and the HII regions to the NE of the NSC, as well as some other features (e.g., the Quintuplet cluster). We note that the light density in the central parsec of the NSC is probably biased -- in total flux and/or in the extinction correction -- due to the low resolution of our data combined with the strong diffuse emission from the mini-spiral and the presence of a few extremely bright sources, such as IRS\,7 or IRS\,1W. We therefore also masked a rectangular area of $1\times1$\,pc centred on Sgr\,A*  Our fits are dominated by large scales, and we did generally not find any significant change in the best-fit parameters when we applied the central masking or not, but occasionally more consistent results were obtained with the central mask  for given image sizes or model combinations. We use the uncertainty maps to apply normal weighting.

All models are elliptical. We therefore use the modified projected radius
\begin{equation}
p = \sqrt{x^2 + (y/q)^2},
\end{equation}
where $x$ and $y$ are the 2D coordinates and $q$ is the ratio between minor and major axis. A King model \citep{King:1962kx}  is then given by 
\begin{equation}
I(x,y) = I_{0, King}  \left[\frac{1}{\sqrt{1+(p/r_{c})^{2}}} - \frac{1}{\sqrt{1+c^{2}}}\right]^{2},
\end{equation}
with $I(x,y)$ the projected surface brightness at a given position, $r_{c}$ the core radius, and $c=r_{tidal}/r_{c}$ the concentration parameter. Leaving $c$ as a free parameter in the fits or fixing it to $r_{tidal}/r_{c} = 15$ as in \citet{Seth:2006uq}, did not result in any significant change of the other best-fit parameters. 

A Moffat-like profile \citep{Moffat:1969zr} is given by
\begin{equation}
I(x,y) = I_{0, Moffat}  \frac{1}{(1+p^{2})^{\beta}}.
\end{equation}`
For the S\'ersic profile  we adopted $b_{n} = 1.9992*n-0.3271$ \citep{Graham:2001ys} and used
\begin{equation}
I(x,y) = I_{e}  \exp{\left\{-b_{n}\left[\left(\frac{p}{R_{e}}\right)^{1/n}-1\right]\right\}}.
\end{equation}
Here, $R_{e}$ is the effective radius, which encloses $50\%$ of the light,  and $I_{e}$ is the surface brightness at $R_{e}$.

The best-fit solution was found via a Levenberg-Markquardt algorithm with the MPFIT software, coded in IDL \citep{Markwardt:2009fk}. Since the residual images do not show any significant systematic effects (except at $p\lesssim1$\,pc, where our data are probably biased and/or our models not adequate, see above and discussion in Section\,\ref{sec:discussion}), we assumed that the models were valid to describe the NSC and rescaled the uncertainties of the best-fit parameters delivered by the Levenberg-Markquardt algorithm  to a reduced $\chi^{2}=1$. These uncertainties will be termed ``formal uncertainties'' in the following in order to distinguish them from the systematic uncertainties that result from our initial assumptions for the model fits and are examined in the following sections.

The MWNSC is not isolated but embedded into the combined emission from NSD, Galactic bulge and Galactic disc (in decreasing order of flux density). Of those, the component with the overall highest flux density and smallest angular scale is the NSD. Its scale length is $>120\,$pc along the Galactic plane and $\sim$$45$\,pc in the vertical direction \citep{Launhardt:2002nx}. We found that the flux offset due to this fore- and background emission is an important source of systematic uncertainty, along with the overall size of the images to which we fit the models.  In the following sections, we describe our model-fitting and  assessment of the relevant systematic uncertainties. For simplicity, we only use the term``background'' when referring to the fore- and background emission into which the MWNSC is embedded. 

\subsection{Constant flux offset}

\begin{table*}
\caption{Ranges of best-fit model parameters for the MWNSC in the IRAC/Spitzer $4.5\,\mu$m image, assuming constant fore-and background light. The first line for each model refers to the PAH and extinction corrected image and  the second line to the corrected and symmetrised image.}
\label{tab:constmodels}
\centering
\begin{tabular}{l c c c c c c c c}
\noalign{\smallskip}
\hline\hline
\noalign{\smallskip \smallskip}
{\bf S\'ersic} & & & & & & & \\
\noalign{\smallskip}
\hline
& $\Delta x$ & $\Delta y$ & $\Theta$ & $I_{e}$ & $q$ & $n$ & $R_{e}$  & $\chi^{2}_{\mathrm red}$\\ 
& (pc)  & (pc) & (deg) & (mJy\,arcsec$^{-2}$) &  & & (pc)  & \\ 
\noalign{\smallskip}
\hline
\noalign{\smallskip}
\noalign{\smallskip}
corrected  & $[0.04,0.05]$ & $ [-0.20,0.03]$ & $[-2.7,-2.8]$ & $[1.5,4.9]$ & $[0.61,0.64]$ & $[1.0,3.5]$ & $[3.8,10.0]$ & $[1.11,1.20]$\\
symmetric & $[0.04,0.04]$ & $ [-0.10,-0.10]$ & $[-0.1,-0.3]$ & $[0.8,4.9]$ & $[0.61,0.63]$ & $[1.2,3.7]$ & $[4.0,13.5]$ & $[1.57,1.64]$\\
\noalign{\smallskip}
\hline\hline
\noalign{\smallskip \smallskip}
{\bf King} & & & &  & & & & \\
\hline
\noalign{\smallskip}
& $\Delta x$ & $\Delta y$ & $\Theta$ & $I_{0}$ & $q$ &  & $r_{c}$ & $\chi^{2}_{\mathrm red}$\\ 
& (pc)  & (pc) & (deg) & (mJy\,arcsec$^{-2}$) &  & & (pc)  & \\ 
\noalign{\smallskip}
\hline
\noalign{\smallskip}
\noalign{\smallskip}
corrected & $[0.05, 0.07]$ & $[-0.26,-0.10]$ & $[-3.4,-1.6]$ & $[16.4,37.6]$ & $[0.59 ,0.65]$ &  & $[1.7,3.2]$ & $[1.11,1.20][$\\
symmetric & $[0.04,0.04]$ & $[-0.10,-0.10]$ & $[-0.3,-0.1]$ & $[16.4,30.6]$ & $[0.60,0.64]$ &  & $[1.9,3.6]$ & $[1.61,1.74]$\\
\noalign{\smallskip}
\hline\hline
\noalign{\smallskip \smallskip}
{\bf Moffat} &  & & & & & & & \\
\hline
\noalign{\smallskip}
& $\Delta x$ & $\Delta y$ & $\Theta$ & $I_{0}$ & $q$ & $\beta$ &  & $\chi^{2}_{\mathrm red}$\\ 
& (pc)  & (pc) & (deg) & (mJy\,arcsec$^{-2}$) &  & &   & \\ 
\noalign{\smallskip}
\hline
\noalign{\smallskip}
\noalign{\smallskip}
corrected & $[-0.03,-0.00]$ & $[-0.05,-0.02]$ & $[-3.6,-1.1]$ & $[341,727]$ & $[0.59,0.62]$ & $[0.72,0.98]$ &   & $[1.22,1.36]$\\
symmetric & $[0.04,0.04]$ & $[-0.10,-0.10]$ & $[-0.1,-0.2]$ & $[359,725]$ & $[0.62,0.63]$ & $[0.71,0.95]$ &   & $[1.69,2.01]$\\
\noalign{\smallskip}
\hline\hline
\end{tabular}
\end{table*}

As a zero-order approximation, we assumed that it is valid to approximate the surrounding background light by a constant offset in flux density \citep[see also][]{Graham:2009lh,Schodel:2011ab}. We fitted all models to quadratic images with sizes of $15\times15$, $20\times20$, and $30\times30$\,pc$^{2}$ and with the flux offset determined in annuli of $4$\,pc width, centred on Sgr\,A* and with inner radii of $10$, $15$, $20$, and $25$\,pc.  The best-fit parameters did not vary strongly with image size ($\lesssim10\%$ in most cases) and did not show any clear systematic trend. The diameter of the annulus used for background subtraction, however, resulted in significant systematic effects for some of the parameters, in particular the half light and core radii, the related light intensity parameters, and the S\'ersic and Moffat indices. This is to be expected if the background light is not constant across the MWNSC's area or if the annulus is too small and thus picks up light from the MWNSC itself. 


We also applied the model fitting to the symmetrised image of the MWNSC (Fig.\,\ref{Fig:symmetry}).  The results of the fits for the different models are summarised in the different sections of Table\,\ref{tab:constmodels} in the rows with the labels ``corrected'' (fit to a fully corrected image) and ``symmetric'' (fit to a fully corrected and symmetrised image). The values in the brackets show the range of the best-fit parameters  obtained for the four different background annuli.  We note that the best-fit centring and rotation angle parameters for the fits to the symmetrised image are different from zero. They thus indicate the accuracy of our method.  

As the numbers in Table\,\ref{tab:constmodels} demonstrate, the assumption of constant background light does not lead to well constrained solutions.  Therefore, to determine meaningful and consistent measurements of the cluster parameters, we must explore models with spatially variable background light.

\begin{figure}[!htb] 
\includegraphics[width=\columnwidth,angle=0]{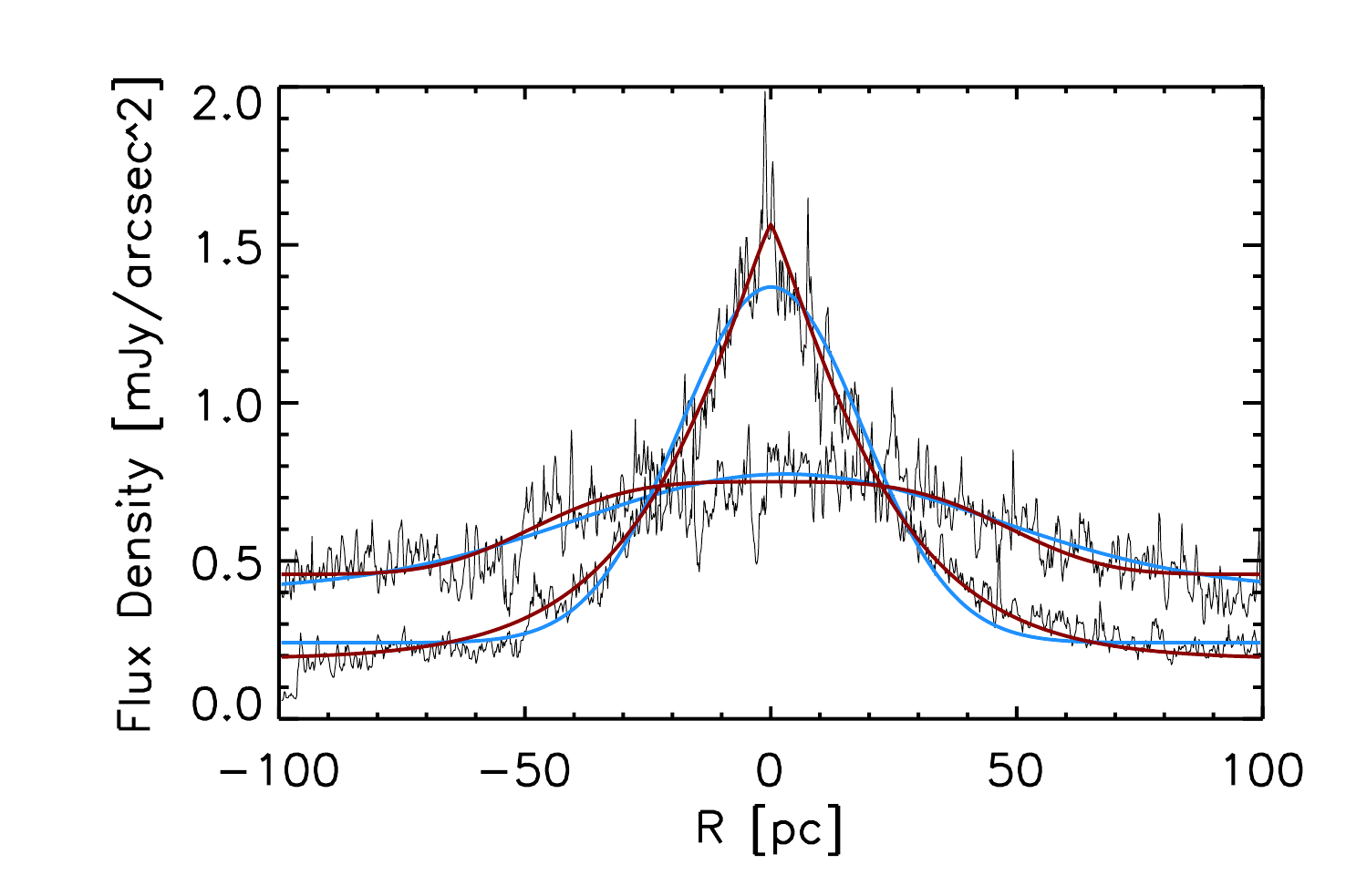} 
\caption{\label{Fig:NSD-profile} Mean flux density profiles (black)  horizontally (broad profile) and vertically (narrow profile)  through the NSD at positions $\sim$20\,pc to the north/south and east/west of Sgr\,A*. The blue lines are best-fit Gaussians and the red lines are best-fit S\'ersic profiles.}
\end{figure}

\subsection{Variable flux offset \label{sec:variable}}

As the preceding paragraph shows,  the assumption of a constant flux for the fore- and background light is probably too simplistic. In fact, the flux density of the NSD is expected to peak at the position of Sgr\,A*. To illustrate this point, we show horizontal and vertical flux density profiles of the NSD near Sgr\,A* in Fig.\,\ref{Fig:NSD-profile}, which were obtained by taking the median of the flux density over corresponding $4$\,pc-wide stripes offset $\sim$ 20\,pc north and south and east and west from Sgr\,A*, respectively. As can be seen, the profiles can be reasonably well fit by Gaussians and the scale length in the vertical direction is significantly smaller than in the horizontal direction. As a result, the underlying NSD may bias the inferred shape of the NSC. S\'ersic profiles provide even better fits, in particular along the vertical direction, where the light density shows a pronounced central peak and where the wings appear to display a profile different from a  Gaussian.

\begin{table*}
\caption{Best-fit model parameters for the variable background light models.}
\label{tab:NSD}
\centering
\begin{tabular}{l l l l l l l l}
\hline\hline
{\bf S\'ersic} & & & & & & \\
\hline
 ID & Size & $I_{e}$ & $q$ & $n$ & $R_{e}$  & $I_{constant}$ & $\chi^{2}_{\mathrm red}$\\ 
 & (pc) & (mJy\,arcsec$^{-2}$) &  & & (pc)  & (mJy\,arcsec$^{-2}$) & \\ 
\hline
1 & $100$  & $0.27$ & $0.37$ & $1.33$ & $84$ & $0.14$ & $1.6$\\
2 & $200$  & $0.28$ & $0.36$ & $1.27$ & $85$ & $0.14$ & $1.8$\\
3 & $100$  & $0.35$ & $0.35$ & $1.3$ & $89$ & $0.20$ & $3.0$\\
4 & $200$  & $0.35$ & $0.32$ & $1.2$ & $96$ & $0.20$ & $3.0$\\
5 & $[60,100]$  & $0.28$ & $0.36$ & $1.1$ & $82$ & $0.14$ & $1.8$\\
\hline\hline
{\bf Gauss} & & & & & & \\
\hline
 ID & Size & $I_{0}$ & $q$ & $FWHM_{major}$ &  $FWHM_{minor}$ & $I_{constant}$ & $\chi^{2}_{\mathrm red}$\\ 
& (pc) & (mJy\,arcsec$^{-2}$) &  & (pc) & (pc)  & (mJy\,arcsec$^{-2}$) & \\ 
\hline
6 & $100$  & $1.07$ & $0.39$ & $39$ & $15$ & $0.26$ & $1.6$\\
7 & $200$  & $0.78$ & $0.38$ & $59$ & $22$ & $0.16$ & $1.9$\\
8 & $100$  & $1.345$ & $0.37$ & $42$ & $15$ & $0.36$ & $3.1$\\
9 & $200$  & $1.021$ & $0.34$ & $64$ & $22$ & $0.24$ & $3.2$\\
\hline\hline
\end{tabular}
\end{table*}

To infer a model of the background light, we masked a rectangular region of $15\times15$\,pc$^{2}$, centred on Sgr\,A*, in the fully corrected $4.5\,\mu$m image and fitted both Gaussian and S\'ersic profiles plus a constant flux offset to the large-scale flux distribution. We centred these models on Sgr\,A* and forced their major axes to be parallel to the GP.  This procedure was applied to images of 100\,pc$\times$100\,pc and 200\,pc$\times$200\,pc size. Prominent structures, such as the Quintuplet cluster or large IR-dark clouds were masked. We refrained from using even bigger images because on even larger scales the spatially variable flux contribution from the  Galactic bulge may become important and thus add an additional complication and because large areas with significant local emission or dark clouds, such as the Sgr B2 region (see Fig.\,\ref{Fig:corrmaps}), would have had to be masked in those larger images. The  best-fit parameters of the S\'ersic and Gaussian background light models are listed in Table\,\ref{tab:NSD} under model ID numbers 1, 2, 6, and 7. The formal uncertainties of the best-fit parameters are far less than their systematic differences for the different image sizes, which is why we do not include the uncertainties in the Table.  We also performed the model fits to the fully corrected and symmetrised image. The results are listed under model ID numbers 3, 4, 8, and 9. Finally, model ID 5 shows the best-fit parameters from a simultaneous double-S\'ersic fit to both the background and MWNSC (see below). We note that the best-fit parameters of the S\'ersic models do not show any strong dependence on the image size, while there is a clear dependence of the best-fit FWHM of the major and minor axes on the  image size for the Gaussian models. All S\'ersic models provide very similar solutions, with S\'ersic indices of $n=1.1-1.3$, flattening parameters of $q=0.32-0.37$, and effective radii of $R_{e}=82-96$\,pc.

\begin{figure}[!thb] \includegraphics[width=\columnwidth,angle=0]{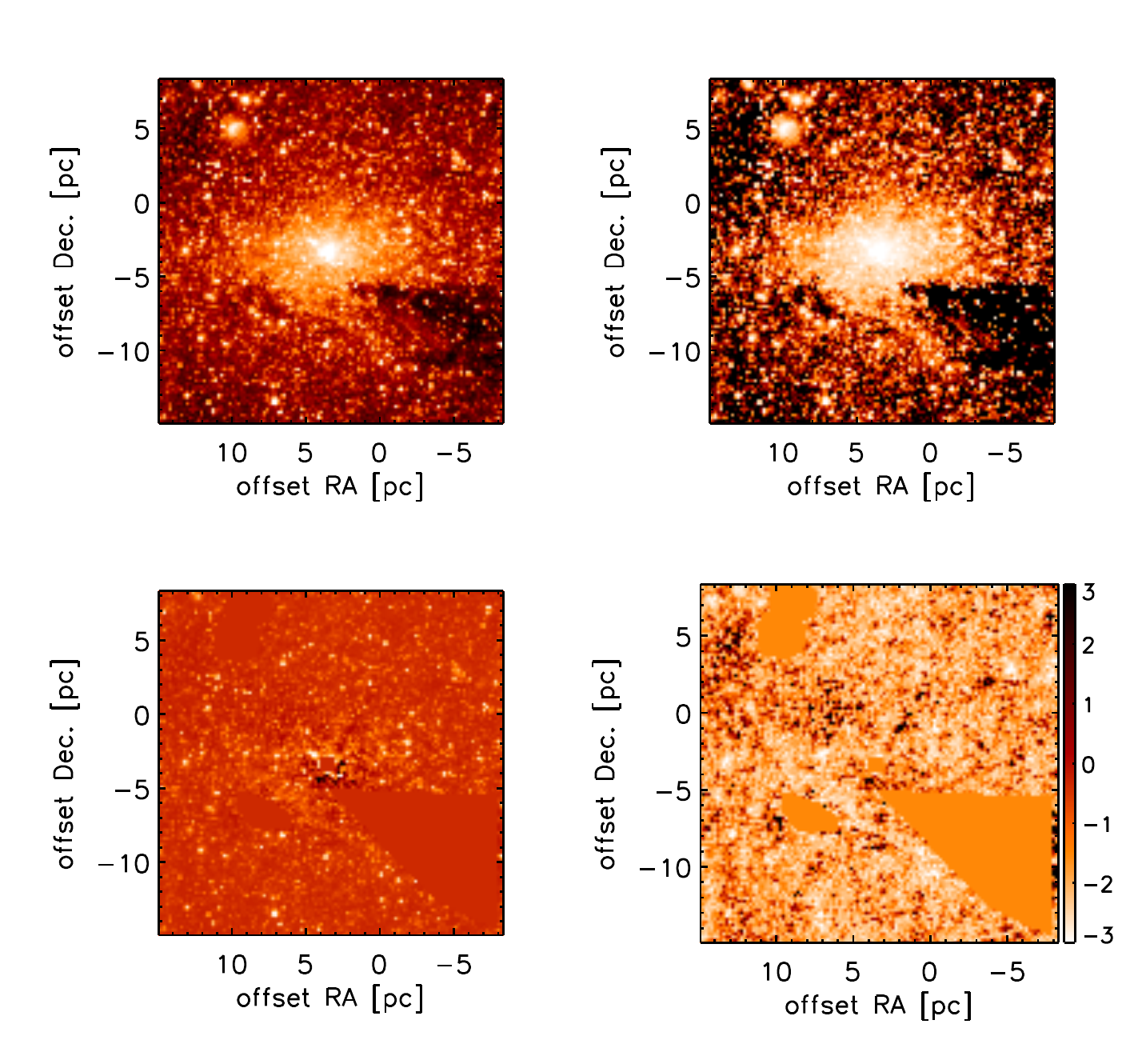} \caption{\label{Fig:resid} Top left: PAH and extinction corrected image of the MWNSC (similar to the one in the right panel of Fig.\,\ref{Fig:NSC}). Top right: Image as in top left, but after subtraction of a S\'ersic model for the background light (model ID 1 in Table\,\ref{tab:NSD}).  Bottom left: Residual of the S\'ersic model fit to the MWNSC after subtracting of the background light. Right: The residual image normalized by the pixel uncertainties.  The colour bar unit is in standard deviations, $\sigma$. The featureless regions in the bottom left and right panels are areas that were masked for the fit. All colour scales are logarithmic.}
\end{figure}

We subtracted these four background light models from the $4.5\,\mu$m image (see, e.g., top right panel in Fig.\,\ref{Fig:resid}) and fitted the three different mathematical models for the  MWNSC, using  image sizes of 16\,pc$\times$16\,pc, 23\,pc$\times$23\,pc, 31\,pc$\times$31\,pc, and 39\,pc$\times$39\,pc. We did not observe any systematic trend of the best-fit parameters with image size and thus quadratically added the uncertainties resulting from the different image sizes to the formal uncertainties of the minimization algorithm. The results are listed in Table\,\ref{tab:models} in the rows labelled as ``Gauss$_{100}$'', ``Gauss$_{200}$'', ``S\'ersic$_{100}$'', and ``S\'ersic$_{200}$''. We performed the same fitting procedure on the symmetrised image of the fully corrected $4.5\,\mu$m emission, with the corresponding results  labelled as ``Gauss$_{100}$, sym.'', ``Gauss$_{200}$, sym.'', ``S\'ersic$_{100}$, sym.'', and ``S\'ersic$_{200}$, sym.''.

We note that the Gaussian background light model leads to significantly different values of some of the best-fit parameters for the MWNSC S\'ersic model, such as the effective radius or the S\'eric index. Also, in case of the MWNSC King model, the assumption of a Gaussian background light model leads to poorly constrained values of the concentration parameter. The fits with the S\'ersic background light model, on the other hand, lead to better constrained and consistent solutions. We interpret this as additional evidence that a S\'ersic model is better suited to modelling the background light. For this reason we only consider solutions with S\'ersic background light models in the following. 
 
Finally, we simultaneously fitted combinations of a S\'ersic law for the background plus the different mathematical models for the MWNSC to the data. The results of the fits for the different models are summarised in the rows of Table\,\ref{tab:models} that follow the  labels ``S\'ersic$_{var}$'', as well as  ``S\'ersic$_{var}$, sym.'' for model fits to the symmetrised image and  ``S\'ersic$_{var}$, diff. sym.'' for model fits to the symmetrised image of the diffuse emission. To assure adequate fitting of both the NSD and NSC, we performed this fit on large images of $86\times86$\,pc$^{2}$, $101\times101$\,pc$^{2}$, $117\times117$\,pc$^{2}$, and $156\times156$\,pc$^{2}$. For the diffuse emission, we could only use smaller fields of $39\times39$\,pc$^{2}$, $59\times59$\,pc$^{2}$, and $78\times78$\,pc$^{2}$, because of the limited size of the point-source-subtracted image (see Section\,\ref{sec:photometry}).  For these combined fits to the NSD and MWNSC we fixed the centres of both S\'ersic models to the position of Sgr\,A* and assumed alignment of both the NSD and the MWNSC with the GP.

We note that the resulting best-fit S\'ersic models for the background light (dominated by the NSD) have flattening parameters of $q\approx0.35$ and S\'ersic indices of $n\approx1.3$ (see Table\,\ref{tab:NSD}). The superposition of the strongly flattened NSD with the MWNSC results in the latter having less ellipticity and smaller half-light radii than when assuming a constant distribution of the background flux.

\subsection{Best-fit models and parameters}
 
\subsubsection{Choice of model.} All three models provide largely equivalent descriptions of the large-scale structure of the NSC. The residual images for all models look very similar. For illustrative purposes we show the unweighted and the weighted residual images for a S\'ersic model for the MWNSC and for the case of the background light described by model ID 1 in Table\,\ref{tab:NSD}.

The reduced $\chi^{2}$ values are similar and range between values of about 1 and 3 in all cases. The Moffat model in general shows slightly higher  reduced $\chi^{2}$ values. It also results in more variable best-fit parameters in between the different fits. In addition, the rotation angle and flattening parameter  show stronger variability and not very good consistency among the different runs and clear deviations from their best-fit values in the case of the S\'ersic and King models. This suggests that the Moffat model is somewhat less suited than the Kind and S\'ersic models to describing the data presented in this paper.

\begin{table*}
\centering
\caption{Best-fit model parameters for the MWNSC in the IRAC/Spitzer $4.5\,\mu$m image with a variable background light model.}
\label{tab:models}
\begin{tabular}{l l l l l l l l l}
\hline\hline
\noalign{\smallskip}
{\bf S\'ersic} & & & & & & & \\
\hline
ID  & $\Delta x$ & $\Delta y$ & $\Theta$ & $I_{e}$ & $q$ & $n$ & $R_{e}$  & $\chi^{2}_{\mathrm red}$\\ 
& (pc)  & (pc) & (deg) & (mJy\,arcsec$^{-2}$) &  & & (pc)  & \\ 
\hline
1 Gauss$_{fixed,100}$\tablefootmark{a} & $0.05\pm0.02$ & $-0.10\pm0.01$ & $-0.0\pm0.5$ & $3.6\pm0.2$ & $0.66\pm0.01$ & $1.9\pm0.1$ & $4.8\pm0.1$ & $2.81\pm0.07$\\
2 Gauss$_{fixed,100}$, sym.\tablefootmark{b} & $0.05\pm0.02$ & $-0.10\pm0.01$ & $-0.0\pm0.4$ & $3.6\pm0.1$ & $0.62\pm0.01$ & $1.7\pm0.1$ & $4.8\pm0.1$ & $1.74\pm0.04$\\
3 Gauss$_{fixed,200}$\tablefootmark{c} & $0.09\pm0.03$ & $-0.19\pm0.04$ & $1.3\pm1.9$ & $1.4\pm0.2$ & $0.63\pm0.02$ & $2.6\pm0.2$ & $9.0\pm0.7$ & $1.32\pm0.08$\\
4 Gauss$_{fixed,200}$, sym.\tablefootmark{d} & $0.04\pm0.02$ & $-0.10\pm0.01$ & $0.7\pm0.6$ & $1.1\pm0.2$ & $0.65\pm0.02$ & $3.5\pm0.4$ & $10.4\pm0.8$ & $2.84\pm0.09$\\
5 S\'ersic$_{fixed,100}$\tablefootmark{e} & $0.07\pm0.03$ & $-0.17\pm0.04$ & $-0.9\pm3.2$ & $4.0\pm0.1$ & $0.72\pm0.02$ & $1.7\pm0.1$ & $3.4\pm0.1$ & $1.32\pm0.08$\\
6  S\'ersic$_{fixed,100}$, sym.\tablefootmark{f} & $0.05\pm0.02$ & $-0.10\pm0.01$ & $-0.3\pm1.0$ & $5.2\pm0.1$ & $0.75\pm0.02$ & $1.6\pm0.1$ & $3.2\pm0.1$ & $2.84\pm0.08$\\
7 S\'ersic$_{fixed,200}$\tablefootmark{g} & $0.06\pm0.03$ & $-0.17\pm0.03$ & $-0.8\pm3.3$ & $3.7\pm0.1$ & $0.72\pm0.02$ & $1.8\pm0.1$ & $3.6\pm0.1$ & $1.32\pm0.08$\\
8 S\'ersic$_{fixed,200}$, sym.\tablefootmark{h} & $0.05\pm0.02$ & $-0.10\pm0.01$ & $0.1\pm1.0$ & $4.4\pm0.1$ & $0.77\pm0.03$ & $2.0\pm0.1$ & $3.6\pm0.1$ & $2.84\pm0.09$\\
9 S\'ersic$_{var}$\tablefootmark{i} & n.a. & n.a. & n.a. & $3.4\pm0.8$ & $0.71\pm0.01$ & $2.0\pm0.1$ & $3.6\pm0.3$ & $1.76\pm0.01$ \\
10 S\'ersic$_{var}$, sym.\tablefootmark{j} & n.a. & n.a. & n.a. & $2.7\pm0.3$ & $0.68\pm0.01$ & $2.4\pm0.3$ & $4.9\pm0.4$ & $1.87\pm0.04$ \\
11 S\'ersic$_{var}$, diff., sym.\tablefootmark{j} & n.a. & n.a. & n.a. & $1.5\pm0.1$ & $0.66\pm0.01$ & $2.8\pm0.2$ & $6.0\pm0.2$ & $0.78\pm0.01$ \\
\hline\hline
\noalign{\smallskip}
{\bf King} & & & &  & & & & \\
\hline
& $\Delta x$ & $\Delta y$ & $\Theta$ & $I_{0}$ & $q$ &  $r_{c}$ & $c$\tablefootmark{k} & $\chi^{2}_{\mathrm red}$\\ 
& (pc)  & (pc) & (deg) & (mJy\,arcsec$^{-2}$) &  & (pc) & (pc)  & \\ 
\hline
1 Gauss$_{fixed,100}$\tablefootmark{a} & $0.04\pm0.02$ & $-0.10\pm0.01$ & $-0.1\pm0.5$ & $28.7\pm1.1$ & $0.65\pm0.01$ & $2.2\pm0.1$ & $21\pm5$ & $2.80\pm0.07$\\
2 Gauss$_{fixed,100}$, sym.\tablefootmark{b} & $0.04\pm0.02$ & $-0.11\pm0.01$ & $-0.1\pm0.4$ & $25.0\pm0.7$ & $0.62\pm0.01$ & $2.4\pm0.1$ & $15\pm2$  & $1.73\pm0.04$\\
3 Gauss$_{fixed,200}$\tablefootmark{c} & $0.04\pm0.02$ & $-0.10\pm0.01$ & $0.9\pm0.6$ & $24.4\pm1.2$ & $0.64\pm0.02$ & $2.3\pm0.2$ & $[42,1427]$ & $2.84\pm0.08$\\
4 Gauss$_{fixed,200}$, sym.\tablefootmark{d} & $0.04\pm0.02$ & $-0.10\pm0.01$ & $0.7\pm0.4$ & $20.9\pm1.1$ & $0.61\pm0.01$ & $2.5\pm0.2$ & $>1000$ & $1.31\pm0.07$\\
5 S\'ersic$_{fixed,100}$\tablefootmark{e} & $0.07\pm0.04$ & $-0.17\pm0.04$ & $-0.3\pm3.5$ & $29.4\pm2.6$ & $0.72\pm0.02$ & $1.7\pm0.1$ & $16\pm3$ & $1.32\pm0.08$\\
6 S\'ersic$_{fixed,100}$, sym.\tablefootmark{f} & $0.05\pm0.02$ & $-0.10\pm0.04$ & $-0.3\pm1.1$ & $33.6\pm1.1$ & $0.76\pm0.02$ & $1.7\pm0.1$ & $14\pm1$ & $2.85\pm0.08$\\
7 S\'ersic$_{fixed,200}$\tablefootmark{g} & $0.07\pm0.04$ & $-0.17\pm0.04$ & $-0.3\pm3.5$ & $28.6\pm2.2$ & $0.72\pm0.02$ & $1.7\pm0.1$ & $18\pm3$ & $1.32\pm0.08$\\
8 S\'ersic$_{fixed,200}$, sym.\tablefootmark{h} & $0.05\pm0.02$ & $-0.10\pm0.02$ & $-0.1\pm1.1$ & $33.8\pm1.2$ & $0.77\pm0.03$ & $1.6\pm0.1$ & $19\pm1$ & $2.84\pm0.09$\\
9 S\'ersic$_{var}$\tablefootmark{i} & n.a. & n.a. & n.a. &  $27.2\pm1.2$ & $0.68\pm0.02$ &  $1.7\pm0.1$ & $35\pm3$ &  $1.76\pm0.01$ \\
10 S\'ersic$_{var}$, sym.\tablefootmark{j} &  n.a. & n.a. & n.a. &  $26.3\pm1.0$ & $0.67\pm0.01$ &  $2.0\pm0.1$ & $25\pm2$ & $1.87\pm0.04$ \\
11 S\'ersic$_{var}$, diff., sym.\tablefootmark{j} &  n.a. & n.a. & n.a. &  $17.6\pm1.4$ & $0.66\pm0.01$ & $2.0\pm0.2$ & $20\pm2$ &  $0.79\pm0.01$ \\
\hline\hline
\noalign{\smallskip}
{\bf Moffat} &  & & & & & & & \\
\hline
& $\Delta x$ & $\Delta y$ & $\Theta$ & $I_{0}$ & $q$ & $\beta$ &  & $\chi^{2}_{\mathrm red}$\\ 
& (pc)  & (pc) & (deg) & (mJy\,arcsec$^{-2}$) &  & &   & \\ 
\hline
1 Gauss$_{fixed,100}$\tablefootmark{a} & $0.03\pm0.01$ & $-0.10\pm0.01$ & $11.4\pm3.4$ & $886\pm189$ & $0.70\pm0.03$ & $0.90\pm0.04$ & & $3.12\pm0.07$\\
2 Gauss$_{fixed,100}$, sym.\tablefootmark{b} & $0.03\pm0.01$ & $-0.10\pm0.01$ & $8.8\pm2.1$ & $906\pm176$ & $0.66\pm0.02$ & $0.91\pm0.04$ & & $2.04\pm0.04$\\
3 Gauss$_{fixed,200}$\tablefootmark{c} & $-0.04\pm0.01$ & $-0.10\pm0.01$ & $4.7\pm1.7$ & $599\pm87$ & $0.65\pm0.02$ & $0.79\pm0.02$ & & $2.98\pm0.07$\\
4 Gauss$_{fixed,200}$, sym.\tablefootmark{d} & $-0.04\pm0.01$ & $-0.10\pm0.01$ & $4.4\pm1.6$ & $562\pm85$ & $0.61\pm0.02$ & $0.79\pm0.02$ & & $1.90\pm0.07$\\
5 S\'ersic$_{fixed,100}$\tablefootmark{e} & $-0.03\pm0.02$ & $-0.05\pm0.02$ & $3.7\pm8.7$ & $790\pm119$ & $0.77\pm0.03$ & $0.99\pm0.04$ & & $1.39\pm0.08$\\
6 S\'ersic$_{fixed,100}$, sym.\tablefootmark{f} & $0.03\pm0.01$ & $-0.11\pm0.01$ & $17.2\pm4.8$ & $855\pm101$ & $0.85\pm0.05$ & $0.99\pm0.03$ & & $3.05\pm0.12$\\
7 S\'ersic$_{fixed,200}$\tablefootmark{g} & $-0.03\pm0.02$ & $-0.05\pm0.02$ & $4.1\pm8.5$ & $759\pm119$ & $0.77\pm0.03$ & $0.97\pm0.04$ & & $1.39\pm0.08$\\
8 S\'ersic$_{fixed,200}$, sym.\tablefootmark{h} & $0.03\pm0.01$ & $-0.11\pm0.01$ & $15.9\pm5.1$ & $771\pm91$ & $0.84\pm0.05$ & $0.95\pm0.03$ & & $3.02\pm0.12$\\
9 S\'ersic$_{var}$\tablefootmark{i} & n.a. & n.a. & n.a. &  $287\pm78$ & $0.53\pm0.01$ & $0.68\pm0.06$ & &   $1.74\pm0.02 $ \\
10 S\'ersic$_{var}$, sym.\tablefootmark{j} &  n.a. & n.a. & n.a. &  $395\pm154$ & $0.54\pm0.03$ & $0.69\pm0.10$ & &   $1.88\pm0.05$ \\
11 S\'ersic, diff., sym.\tablefootmark{j} &  n.a. & n.a. & n.a. &  $365\pm49$ & $0.63\pm0.03$ & $0.79\pm0.02$ & &   $0.81\pm0.01$ \\
\hline
\end{tabular}
\tablefoot{
\tablefoottext{a}{Background light contribution fixed and described by Gaussian as given by ID\,6 in Table\,\ref{tab:NSD}.}
\tablefoottext{b}{Background light contribution fixed and described by Gaussian as given by ID\,8 in Table\,\ref{tab:NSD}.}
\tablefoottext{c}{Background light contribution fixed and described by Gaussian as given by ID\,7 in Table\,\ref{tab:NSD}.}
\tablefoottext{d}{Background light contribution fixed and described by Gaussian as given by ID\,9 in Table\,\ref{tab:NSD}.}
\tablefoottext{e}{Background light contribution fixed and described by S\'ersic model as given by ID\,1 in Table\,\ref{tab:NSD}.}
\tablefoottext{f}{Background light contribution fixed and described by S\'ersic model as given by ID\,3 in Table\,\ref{tab:NSD}.}
\tablefoottext{g}{Background light contribution fixed and described by S\'ersic model as given by ID\,2 in Table\,\ref{tab:NSD}.}
\tablefoottext{h}{Background light contribution fixed and described by S\'ersic model as given by ID\,4 in Table\,\ref{tab:NSD}.}
\tablefoottext{i}{Background light contribution described by  a S\'ersic model with freely variable parameters during the fit. The best-fit parameters are listed under ID\,5 in Table\,\ref{tab:NSD}.}
\tablefoottext{j}{Background light contribution described by  a S\'ersic model with freely variable parameters during the fit.}
\tablefoottext{k}{Brackets indicate a range of best-fit parameters for those cases where the solutions differed strongly between different image sizes.}
}
\end{table*}

\subsubsection{Centring of the MWNSC.} The centre of the cluster coincides with the position of Sgr\,A* in all cases to within  $<1$\,pixel ($5"$ or $0.20$\,pc). Although there appears to be a small systematic average offset of the cluster to the south and east of Sgr\,A*, we do not think that this is a real signal because systematic uncertainties in the extinction correction may introduce some bias, because the flux distribution in the central parsec is probably strongly biased (see above), and because we did not take care to provide high-precision astrometry. In fact, the MWNSC in the symmetrised images should be precisely centred on Sgr\,A*, but also shows some small offset, which can be taken as an indication of our overall astrometric accuracy, which is of the order of $0.1$\,pc.  We therefore conclude that the MWNSC is centred on Sgr\,A* within the limits imposed by the accuracy of our data and models. From star counts in the central parsec we know in fact that the cluster centre coincides with the position of Sgr\,A* to within $<1"$ \citep{Schodel:2007tw}. 

The MWNSC's major axis is remarkably well aligned with the Galactic plane. If we ignore the fits with the Moffat model, which we deem hardly reliable for this parameter (see discussion above), the fits indicate that any possible misalignment between the major axis of the MWNSC and the GP is close to $0^{\circ}$. We therefore conclude that the major axis of the MWNSC is aligned with the GP. 

\subsubsection{Flattening, size, profile, and luminosity.} When computing mean values for some of the best fit parameters in the following, we focus
exclusively on the models with IDs 5-11 in Table \,\ref{tab:models}. We do not take the fits with constant or Gaussian profile of the background light  into account because we think that these cases are less reliable, as discussed above. All mean best-fit parameters are calculated from unweighted fits, and the cited $1\,\sigma$ uncertainties are taken from the error of the mean.

The mean ratio between the minor and the major axes is $q_{MWNSC}=0.71\pm0.02$ for both the King and S\'ersic models, while it is $0.70\pm0.05$ for the Moffat models. We point out that the individual best-fit values for $q$ are not very consistent among the different fits for the Moffat model and deviate strongly from the mean value derived from the other two models. Also, the models fit with a variable background component tend to result in somewhat lower, but not significantly different, values of $q$. We  adopt $q_{MWNSC}=0.71\pm0.02$ as our best estimate of the flattening parameter of the MWNSC.

The  mean of the effective/half light  radii  of the S\'ersic  models is $<R_{h, Sersic}>=(4.0\pm0.4)$\,pc.  Using the term ``half light radius'' implies assuming spherical symmetry, which is not strictly true here. Nevertheless, we cite $R_{h}$ because it is a convenient and frequently used value. We calculated the $R_{h}$ for the King models by assuming a spherical cluster with the core radii, flux densities, and concentration parameters of the respective best-fit parameters. The half-light radii of the King models are roughly $2.3$ times larger than the core radii, and their mean is $<R_{h, King}>=(4.3\pm0.3)$\,pc, which is consistent with the corresponding value for the S\'ersic models. We therefore adopt a mean value of $<R_{h, MWNSC}>=(4.2\pm0.4)$\,pc, the mean from the King and S\'ersic fits, as our best estimate of the half-light radius of the MWNSC. We note that the half-light radius of the best-fit S\'ersic model to the diffuse light appears to be an outlier, but is still marginally consistent with this best estimate on the $\sim$$3\,\sigma$ level. Our best estimate of  $<R_{h, MWNSC}> $ is close to the value of $R_{h}= 4.4$\,pc determined recently by \citet{Fritz:2013hc}, who used  two S\'ersic profiles to fit the MWNSC plus surrounding light. They do not provide any uncertainty for their value of $R_{h}$. 

The mean S\'ersic index is $<n_{MWNSC}>=2.0\pm0.2$. The S\'ersic index of the NSC found here is thus smaller than the one found by \citet{Graham:2009lh} or \citet{Schodel:2011ab}, who found $n=3$ and $n=3.4$, respectively, but higher than the one derived by \citet{Fritz:2013hc} ($n=1.42\pm0.03$). These other studies used data at a different wavelength (K-band, $\sim$$2.2\,\mu$m) did not take the strong differential extinction across the NSC into account, approximated the NSC as spherical and/or had to symmetrise the images to deal with the strong differential extinction, or used different approaches in dealing with the emission from the NSD. It appears that determining the  S\'ersic index can be subject to considerable systematic uncertainties. The same may be true for extragalactic NSCs and should be kept in mind as a caveat when dealing with this parameter. 

We note that all models describe similar power law slopes  at high values of the projected radius, $p$. The power law exponent approaches $2.0$ for the King model outside the core radius, as it does for the for the Moffat model, with $\beta=0.86\pm0.05$ close to $1$. For the mean S\'ersic model, we obtain a projected power law slope  of  $2.0$ at $p=5.0$\,pc \citep[see appendix A1 in][for the formula used to calculate the power law slope for the S\'ersic model]{Graham:2009lh}. This is in good agreement with the results of \citet{Launhardt:2002nx}, who report a similarly steep power law of the cluster profile beyond $p\approx6$\,pc, with the 3D stellar density decreasing proportionally to $r^{-3}$. On the other hand, close to Sgr\,A*, the MWNSC displays the familiar approximate $r^{-1.8}$ 3D-density law \citep[see discussion in][]{Schodel:2007tw} that was already derived from the first NIR observations by \citet{Becklin:1968nx}. We checked this in our data by fitting the radial light profiles of our corrected images (see section\,\ref{sec:structure} and Fig.\,\ref{Fig:profiles}) with power laws for projected radii between 1 and 3\,pc. The projected light density is proportional to $p^{-0.9}$ parallel and perpendicular to the GP, without any significant dependence on whether the background light (using a S\'ersic model, see section\,\ref{sec:variable}) is subtracted or not. 

The mean total luminosity from the S\'ersic and King models is $L_{MWNSC,4.5\,{\mu}m, Sersic}=(4.1\pm0.3)\times10^{7}$\,L$_{\odot, 4.5\,{\mu}m}$ and  $L_{MWNSC,4.5\,{\mu}m, King}=(4.1\pm0.3)\times10^{7}$\,L$_{\odot, 4.5\,{\mu}m}$, respectively, assuming an absolute magnitude of $3.27$ for the Sun in the Spitzer $4.5\,\mu$m band \citep{Oh:2008fk} and a distance modulus of $14.51$. We have included the fits to the diffuse emission, which give similar total luminosities to the fits for the other cases. This is not necessarily surprising because the sigma-filtering that we applied to all rebinned images (see section\,\ref{sec:extinction}) will have largely suppressed the contribution from individual bright stars in all cases.  The luminosities from the S\'ersic and King models are consistent with each other, and we adopt as our best estimate $L_{MWNSC,4.5\,{\mu}m}=(4.1\pm0.4)\times10^{7}$\,L$_{\odot, 4.5\,{\mu}m}$.

\section{MGE fit}
\label{sec:mge}


To compute a surface brightness profile we use the  \emph{$MGE\_\,FIT\_\,SECTORS$} package written by \citet{Cappellari:2002qf}. 
 This set of IDL routines does photometric measurements directly on images  along sectors. The measurements of four quadrants are averaged  under the assumption of point symmetry. Measurements are taken along elliptical annuli with constant axial ellipticity $\epsilon=1-b/a=1-q$.  We chose sectors of five degrees width. On the measurements a multi-Gaussian expansion \citep[MGE,][]{Emsellem:1994ve} fit is performed. Every Gaussian of the series is fully determined by the maximum intensity  $I_j$, standard deviation $\sigma_j$, and the axial ratio $q_j$. Parametrising the surface brightness profile with a series of Gaussian functions has the  advantage that deprojection can be done analytically, resulting in a Gaussian series as well. The influence of every Gaussian on the profile is locally limited. At small radii ($R \ll \sigma$), the Gaussian contributes only a constant value; at large radii ($R\gg\sigma$) the Gaussian decreases rapidly to zero. 
 
\begin{figure}
\resizebox{\hsize}{!}{\includegraphics{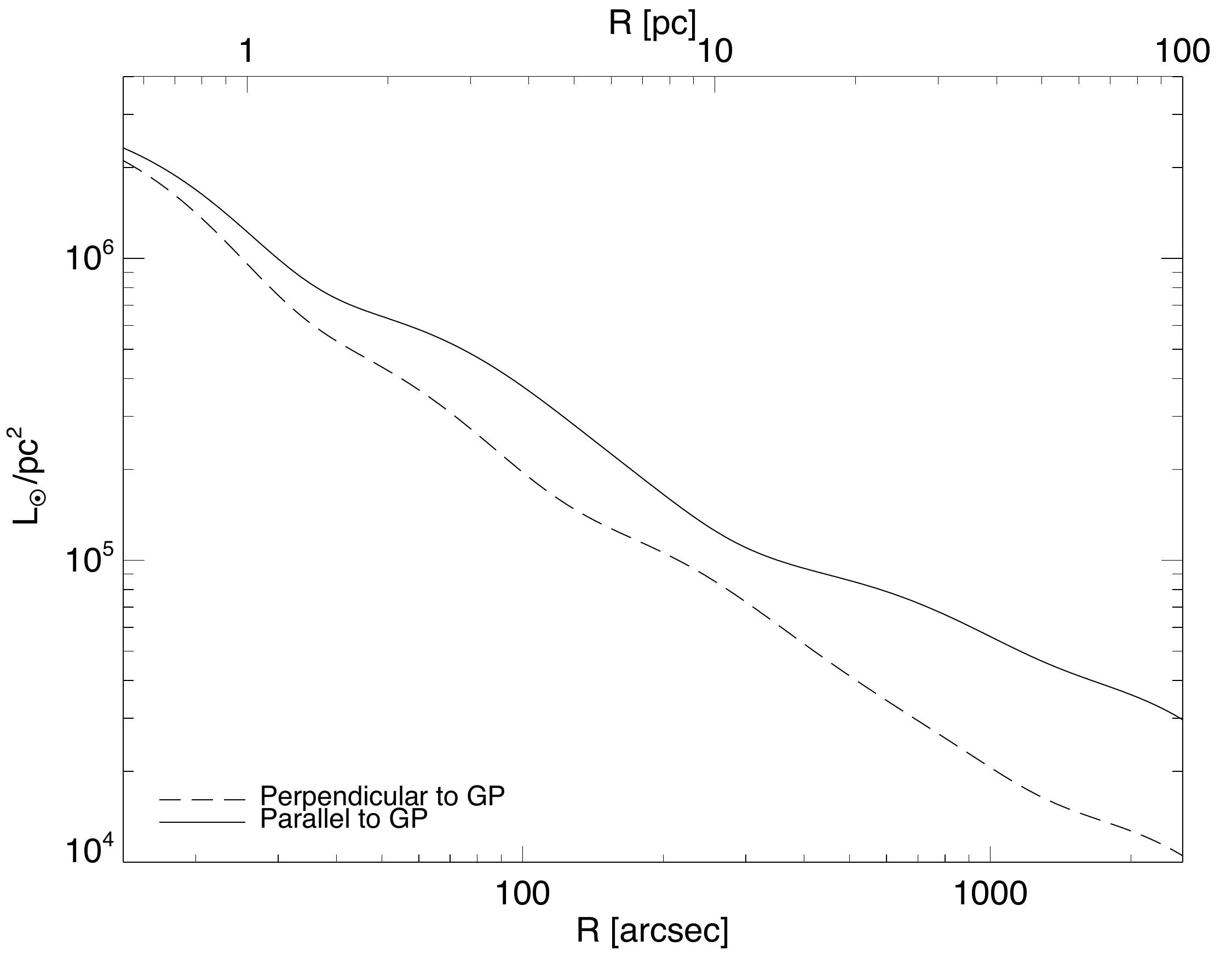} }
\caption{Surface brightness profile derived from an MGE fit to the PAH- and extinction-corrected Spitzer/IRAC 4.5\,$\mu m$ image. The solid line denotes the major axis along the Galactic plane, the dashed line denotes the minor axis perpendicular to the major axis.}
\label{fig:profile_mge}
\end{figure}

The centre of the image is defined as the position of Sgr\,A*, and the photometry is measured along 19 sectors. We masked the 20 km s$^{-1}$ cloud, the Quintuplet cluster, and the central parsec, which may be dominated by emission from the mini-spiral (see preceding section).  The output of the MGE fit is listed in the first three columns of Table \ref{tab:sbprof} for the PAH- and extinction-corrected image. To estimate the uncertainty of the surface brightness profile, we used the uncertainty map and repeated the photometric measurement and MGE fit on the corrected image $\pm$ the uncertainty map. The surface brightness profiles along the major and minor axis are shown in Fig.\,\ref{fig:profile_mge}.
map.   The total luminosity, obtained by summing the contributions from all Gaussians out to $\sigma$ \textgreater 100\,pc, is ($1.6 \pm0.5) \times 10^9\,L_{\odot,4.5\mu m}$.

\begin{table}
\caption{MGE fit parameters for the Spitzer/IRAC PAH- and extinction-corrected image.}
\label{tab:sbprof}
\centering
\begin{tabular}{c c c}
\noalign{\smallskip}
\hline\hline
\noalign{\smallskip}
$I$ &$\sigma$ &$ q$ \\ 
($10^6 L_{\odot,4.5\mu m} / pc^2$) &  (arcsec)   & \\
\noalign{\smallskip}
\hline
\noalign{\smallskip}
 $   25.13  $&$    14.1  $&$   0.90$ \\
 $     2.63  $&$    52.8  $&$   1.00$ \\
 $     1.35  $&$    56.1  $&$   0.35$ \\
 $     2.33  $&$   101.6 $&$   0.38$ \\
 $     0.71  $&$   150.5 $&$   1.00$ \\
 $     0.15  $&$  481.3  $&$   0.09$ \\
 $     0.44  $&$  581.3  $&$   0.42$ \\
 $     0.17  $&$  2656.5$&$   1.00$ \\
 $     0.31  $&$  2656.5$&$   0.20$ \\
\noalign{\smallskip}
\hline
\noalign{\smallskip}
\end{tabular}
\end{table}

The MGE fit is not designed for a decomposition of the  light profile in  contributions from the NSC, the NSD, and bulge. Nevertheless, we note: (a) The profiles parallel and perpendicular to the GP are markedly different, even at distances of $1-2$\,pc from Sgr\,A*. (b) The profile parallel to the GP shows an upward inflection at a projected radius around $10$\,pc, which we interpret as the distance where the additional emission from the NSC stars becomes appreciable over the emission from the NSD, which has a significantly greater scale length along the GP. A corresponding change in slope in the direction vertical to the GP may be present at $\sim$5\,pc. We interpret this difference as being related to both the flattening of the MWNSC along the GP and to the smaller vertical  scale length of the NSD. (c) The mean value of the flattening parameter of the innermost five components of the MGE fit is $<q_{MGE}>=0.73$, consistent with the result of the model fits in the preceding  section.

To give an approximation of  the  total luminosity of the MWNSC, we integrated the luminosity of the five innermost Gaussian components.This is motivated by the fact that the fifth component has a FWHM of $\sim$13\,pc, while the FWHM of the sixth component is already  $\sim$32\,pc, indicating that it obviously traces the larger structure of the NSD.  The integrated luminosity of the five smallest Gaussian components is $L_{MWNSC,4.5\,{\mu}m,MGE} =3.6\,\times\,10^7\,L_{\odot,4.5\mu m}$, in good agreement with the total luminosity of the MWNSC inferred in the preceding section.

\section{Discussion \label{sec:discussion}}

Previous measurements of the overall properties of the MWNSC suffered from the strong differential extinction in the Galactic Centre region that exceeds several magnitudes even at wavelengths of $\lambda\sim2\,\mu$m. Those NIR studies could therefore not provide unambiguous answers to questions such as the extent and symmetry of the cluster, even when they tried to compensate for the differential extinction \citep[e.g.,][]{Catchpole:1990rz,Philipp:1999nx}. In addition, many studies, in particular those at high angular resolution, were limited to regions far inside the half-light radius of the NSC \citep[e.g.,][]{Scoville:2003la,Eckart:1993zr,Genzel:2003it,Schodel:2007tw}.  In this work, we provide a new description of the NSC, based on extinction-corrected MIR data from the Spitzer IRAC survey that overcomes some of the major difficulties and provides an overall view of the NSC at the centre of the Milky Way. In this section we summarise and discuss our main results.

\subsection{Properties of the MWNSC}

Strong, differential extinction has so far impeded an accurate assessment of one of the most basic properties of the MWNSC, which is whether it is intrinsically spherically symmetric or not. By using MIR images, where interstellar extinction is at a minimum, combined with extinction correction, we minimise the influence of extinction on the observed shape of the cluster. The resulting light profiles of the MWNSC are clearly different in directions perpendicular and parallel to the Galactic plane. We further provide evidence for the intrinsic (projected) shape of the MWNSC by a comparison between the original (PAH and extinction corrected)  image and a symmetrised image of the GC at $4.5\,\mu$m. We do not find any significant differences between the two images, except for what can be attributed to the influence of a large-scale IR-dark cloud. We therefore feel that the evidence supports clearly the notion of a flattened MWNSC, symmetric with respect to the Galactic plane and a perpendicular axis through its centre. Such a projected form is in agreement with an intrinsically  flattened, axisymmetric cluster.

While it had been shown previously that stellar number counts in the central parsec peak within $1"/0.04\,$pc of the black hole \citep{Genzel:2003it,Schodel:2007tw}, it was not clear whether the {\it entire} NSC would be centred on Sgr\,A* on {\it large} scales.
Our analysis shows that the cluster centre lies within $\lesssim0.20$\,pc of Sgr\,A*, not more than one pixel of the rebinned images used here.  Considering the low-angular resolution of the images used and the possible systematic uncertainties in the central parsec owing to the presence of strong PAH and /or warm dust emission and a small number of very bright MIR sources, we consider that, on large scales, there is no evidence of any offset between the centre of the NSC and Sgr\,A*. This agrees with the centring of the cluster on  scales of a few 10 mpc that has been found in high angular resolution studies \citep[e.g.,]{Eckart:1993zr,Ghez:1998ad,Genzel:2003it,Schodel:2007tw}. 

The alignment of the MWNSC with the Galactic plane appears to be almost perfect. In combination with its flattening, this is consistent with its rotation parallel to overall Galactic rotation. We find a well constrained value of the ratio of minor to major axis of $q=0.71\pm0.02$, corresponding to an ellipticity $\epsilon=1-q=0.29\pm0.02$. In this point we disagree with the recent work of \citet{Fritz:2013hc}, who claim a spherical shape for the MWNSC and interpret the flattened density contours as being caused by overlap with the NSD. In our work, the flattening of the MWNSC already becomes apparent in the light profiles at $\gtrsim1$\,pc projected distance from Sgr\,A* (see Figs.\,\ref{Fig:profiles}, \ref{Fig:diffuse} and \ref{fig:profile_mge}). Our simultaneous fits of2D models of the MWNSC and the NSD to the data also result in a clearly flattened MWNSC. We interpret our findings as evidence that the flattening of the MWNSC is {\it intrinsic} and not just an apparent effect because of a superposition with the NSD.  The main difference between our work and the work of \citet{Fritz:2013hc} is that we use MIR imaging while they base their analysis on NIR imaging. The NIR image in the left-hand panel of Fig.\,\ref{Fig:NSC_Ks_4.5}, the $3.6\,\mu$m images  in Figs.\,\ref{Fig:maps} and \ref{Fig:colorcheck} and the extinction map in Fig.\,\ref{Fig:ext} show that there is considerable extinction present at offsets $\pm10$\,pc running parallel to the GP toward the north and south, with typical values of $A_{4.5}\approx1$ and thus $A_{1.7}\approx4.4$ \bibpunct[ ]{(}{)}{,}{a}{}{;} \citep[using the extinction law of][ for the wavelength conversion]{Nishiyama:2009oj}. \bibpunct[ ]{(}{)}{,}{a}{}{;}  

At such high values there is the danger of strong systematic uncertainties  in the NIR because there will be little signal in the highly extincted regions. Few stars will be detected in star counts and those will lie preferably in front of the dark clouds. It appears therefore plausible that, in the NIR, extinction correction, masking and symmetrising may suffer from significant systematic effects. Thus, extinction would naturally lead to finding a more flattened NSD around the MWNSC. This, in turn, would make the MWNSC appear less flattened in a double-component fit. The difference between our results and those presented by \citet{Fritz:2013hc} demonstrate that the GC is once more a challenging target and that caution is required when interpreting the data. It is possible that there are still significant systematic effects also  present in the analysis presented in this work. For example, the small uncertainty of the estimated flattening of the MWNSC quoted in this work may underestimate the true uncertainty. In any case,  it would be surprising to find a spherical nuclear star cluster embedded in a strongly flattened NSD, and the rotation of the MWNSC would also be difficult to reconcile with a spherical system. 

As for the density function of the MWNSC, it appears to be described well overall by the $\rho\propto r^{-1.8}$ density law that has been 
established by many studies in the past decades \citep[see, e.g., discussion in][]{Schodel:2007tw}. Here we note that this density law is an approximation for a spherical cluster and only valid within a few parsecs of Sgr\,A*. In fact, the density law is steeper in the direction perpendicular to the GP and flatter along the GP. At distances beyond $p\approx5$\,pc, the cluster profile appears to be significantly steeper \citep[see also][]{Launhardt:2002nx}. There has been much discussion about the existence or not of a stellar cusp around Sgr\,A* in the past years, triggered by the finding of a flat core of old stars within about $0.5$\,pc of the MBH (see Sect.\,\ref{sec:dynamics}). Here we add two minor remarks to the discussion: (1) The stellar cusp will form within the influence radius of the MBH, which contains roughly the stellar mass corresponding to the mass of the BH. The radius of influence lies between 1\,pc and 3\,pc in case of the GC \citep[e.g.,][]{Trippe:2008it,Schodel:2009zr,Merritt:2010ve}. The observed $\sim$$r^{-1.8}$ density law at these distances from Sgr\,A* agrees well with the canonical density law of a stellar cusp \citep[$\rho\propto r^{-1.75}$, e.g.,][]{Bahcall:1977ys,Lightman:1977ly,Murphy:1991zr,Preto:2010kx}. The ``missing cusp'' problem at the GC only refers to the region within $\sim$0.5\,pc of Sgr\,A*. (2) It would be of interest to investigate the density law of a stellar cusp in a {\it rotating} stellar system.

We derive a total luminosity for the MWNSC of $L_{NSC,4.5\,\mu{m}}=(4.1\pm0.4)\times10^{7}\,L_{\odot}$. Lower bounds to this value are provided from the innermost five components of the  MGE fitting ($L_{NSC,4.5\,\mu{m}}=3.6\times10^{7}\,L_{\odot}$). Recent research into the stellar mass-to-luminosity ratio in the MIR has come to the result that it is largely constant, i.e.\ independent of the properties and history of the stellar population. \citet{Meidt:2014fk} find $\Upsilon_{\ast}^{3.6\mu{m}}=0.6\pm0.1$.  According to the modelling of MIR mass-to-light ratios by \citet{Oh:2008fk} we can use the same value at $4.5\,\mu$m because the uncertainty of the simple wavelength conversion from $3.6\,\mu$m to $4.5\,\mu$m will be significantly smaller than the uncertainty of $\Upsilon_{\ast}^{3.6\mu{m}}$. We assume $\Upsilon_{\ast}^{4.5\mu{m}}=0.6\pm0.1$, therefore, which also agrees with the value  $\Upsilon_{\ast}^{4.5\mu{m}}=0.6\pm0.2$ determined from modelling of spectroscopic data of the MWNSC by Feldmeier et al. (submitted to A\&A). 

We thus derive a mass of $M_{MWNSC}=(2.5\pm0.2_{stat}\pm0.4_{syst})\times10^{7}$\,M$_{\odot}$, where the systematic uncertainty reflects the uncertainty of $\Upsilon_{\ast}^{4.5\mu{m}}$. This value lies between the $M_{NSC}=3.0\pm1.5\times10^{7}$\,M$_{\odot}$ derived by \citet{Launhardt:2002nx} and the $(1.3\pm0.3)\times10^{7}$\,M$_{\odot}$ estimated by \citet{Fritz:2013hc} based on isotropical spherical Jeans modelling (for a GC distance of 8\,kpc) and agrees with both on the $1\,\sigma$ level. We note that our measurements are based on a completely different data set and at a different wavelength than the work of \citet{Launhardt:2002nx}, who used the K-band measurements of \citet{Philipp:1999nx}. In particular, our measurements are significantly less affected by extinction, take the non-spherical shape of the NSC into account, and profit from the low uncertainty of the stellar mass-to-light ratio in the MIR. The complementarity of the data and the reduced uncertainty in our work give us confidence in the accuracy of the mass measurement of the NSC, which thus contains about five times as much mass as the central black hole, Sgr\,A*.

We note that the models fitted to the MWNSC images in this work are optimised to describe its overall properties on large scales and at relatively low linear resolution and would like to remind the reader that it is  well established that the MWNSC has a core radius of $\sim$$0.25-0.4$\,pc \citep[e.g.,][]{Schodel:2007tw,Buchholz:2009fk}. This corresponds roughly to the central $2\times2$ pixels in our rebinned map. Care should therefore be taken when using our results for modelling the inner parsecs of the MWNSC. A different approach, e.g., with a broken power law, should then be chosen \citep[see][]{Do:2009tg,Schodel:2009zr}. On the other hand, the overall properties of the MWNSC do not seem to be affected significantly  by the exact choice of the model. This is demonstrated by the agreement of the best-fit angles, half light radii, flattening parameters, and total luminosities between the King and S\'ersic models.

Overall, the main characteristics of the MWNSC, i.e.\ its half light radius and luminosity/mass agree well with the properties of extragalactic NSCs. The luminosity/mass of the MWNSC lies at the higher end of the observed values, but is not unusual considering that the Milky Way is a relatively massive galaxy \citep[see, e.g., Fig.\,14 in ][]{Rossa:2006zr}.

\subsection{Implications for NSC formation}

The strong flattening of the MWNSC, as well as its alignment with and rotation parallel to the Galactic disc, agrees with the findings of \citet{Seth:2006uq} for extragalactic NSCs in edge-on spirals and supports their model of repeated in situ formation of stars in accreted gas discs. In fact, a disc of young stars is observed in the GC right now \citep[e.g.,][]{Levin:2003kx,Paumard:2006xd,Lu:2009bl}.  One of the problems of the model discussed by \citet{Seth:2006uq} is how subsequent star formation in disc components from gas infalling from the galaxy plane can form an ellipsoid/spherical system. The case of the Milky Way shows that discs of recently formed stars must not necessarily be aligned with the galaxy plane. In fact, the clockwise system of stars in the central parsec lies at an angle of roughly 60\,degrees with the Galactic plane \citep{Paumard:2006xd}. As a result, while infalling gas will be aligned with the galaxy disc {\it on average}, this must not be true for an individual event. This is not surprising, given that the vertical extent of the NSC is orders of magnitude smaller than the scale height of the Galactic disc. About 50\% of the few million-year-old stars in the central parsec of the GC do not form part of the clockwise disc and appear to be distributed in a more isotropic way - or form part of a less-well defined disc/streamer \citep{Bartko:2009fq,Lu:2009bl}. This provides further evidence that in situ star formation from infalling gas may be able to create spheroidal systems. The flattening and rotation of the MWNSC support the notion that infall of material occurs, on average, along the GP. This means that a certain connection exists between the Galactic disc and the MWNSC.

\citet{Antonini:2012fk} have studied the formation of the MWNSC by the repeated infall and merger of globular clusters. Their simulations can result in a flattened cluster with an axis ratio close to the one measured in this work. However, this result only holds for the inner 10\,pc of their simulated cluster, which, in addition, has a half-mass radius about twice as large as the one measured here for the MWNSC (assuming a constant mass-to-light ratio). Also, they find a low degree of rotation, which may contradict observations that indicate that the rotation of the NSC at radii of a few parsecs is of the same order of magnitude as its velocity dispersion \citep{Trippe:2008it,Schodel:2009zr}. It must be pointed out that models are usually fine-tuned to reproduce the current state-of-the art of observational knowledge. It is therefore possible that the infall of globular clusters may have provided a significant contribution to the stellar population of the MWNSC. No observational evidence still exists for the globular cluster infall scenario. The in situ formation scenario, however, is clearly supported by observations. Finally, \citet{Hartmann:2011uq} have compared integral-field spectroscopic observations of nuclear clusters with dynamical models and conclude that purely stellar dynamical mergers cannot reproduce the observations. On the other hand, they also exclude a formation scenario based on gas infall and only in situ formation. It is therefore likely that both processes contribute to the formation of NSCs.

\subsection{Implications for stellar dynamics \label{sec:dynamics}}

Our improved knowledge of the overall properties of the MWNSC are fundamental to understanding its formation and future evolution, as well as to interpreting observations of external systems that suffer from linear resolutions orders of magnitude smaller than in the Milky Way. A question of stellar dynamics that has attracted considerable attention in the past years is the problem of the formation of a stellar cusp around the central black hole. While cusp formation is a firm prediction of theoretical dynamics \citep[e.g.,][]{Alexander:2006fk,Merritt:2013uq,Preto:2010kx}, the distribution of the old -- and therefore dynamically relaxed -- stars within $0.5\,$pc of Sgr\,A* is significantly flatter than predicted by theory \citep{Buchholz:2009fk,Do:2009tg,Bartko:2010fk,Do:2013fk}. Several ideas have been forwarded to explain this discrepancy between theory and observations, among them that collisions may destroy the envelopes of giant stars and thus render the cusp invisible \citep[e.g.,][]{Dale:2009ca,Amaro-Seoane:2014fk} or that the cluster formed with a large core and has not yet reached equilibrium \citep{Merritt:2010ve}. 

What are the implications of the axisymmetry that we find here for the MWNSC?  Deviation from sphericity has been addressed in the context of triaxial bulges, bars, or stellar discs on scales of 100\,--\,1000~pc, but also a number of theoretical studies have investigated non-spherical structures of the nucleus itself \citep{Holley-Bockelmann:2001fk,Holley-Bockelmann:2002uq,Berczik:2006kx,Merritt:2011cr,Vasiliev:2013ys,Khan:2013zr}.  The origin of the non-sphericity in these studies can be the merger with another nucleus \citep{Milosavljevic:2001ly} or dissipative interactions between the stars and a dense accretion disc \citep{Rauch:1995ve}.  Deviations from spherical symmetry are important in the study of galactic nuclei for two reasons.

(i) One is the potential \emph{temporary} boost in disruption rates of extended stars or in the capture via gravitational radiation of compact ones \citep{Poon:2001qf,Holley-Bockelmann:2001fk,Holley-Bockelmann:2002uq,Merritt:2004bh,Poon:2004dq,Merritt:2011cr}.  As discussed in \citet{Amaro-Seoane:2012nx}, deviations from non-sphericity lead to orbits that can get very close to the centre, the so-called ``centrophilic'' orbits. At distances within the sphere of influence, a significant percentage of stars might be on centrophilic orbits.  The reason we call them temporary is that this would lead to a consequent depletion of stars in the loss cone, with the implication that current rates would drop, although this depends on the lifetime of the deviation from sphericity.  The problem is not an easy one to model, so we usually have to resort to large simplifications.  In particular, we must explore the behaviour of the potential \emph{very} close to the MBH because, by definition,  the potential is completely dominated by the MBH at some point and, thus, spherically symmetric.  The implications are still debated. Recently, \citet{Vasiliev:2013oq} have performed statistical models calibrated with direct $N-$body simulations for different values of the capture radius and the amount of flattening and found that the rates are only slightly enhanced, by a factor of a few.

(ii) The second reason is the driving of massive binaries of black holes to distances below one parsec, the so-called ''last parsec problem'', in nuclei without gas.  It has been claimed with direct-summation $N-$body integrations of galactic nuclei with an initial amount of rotation that triaxiality or axisymmetry alone drives the binary efficiently to coalescence in less than a Hubble time \citep{Berczik:2006kx}.  However, \citet{Vasiliev:2013kl} have recently shown, also with direct-summation simulations, that the shrinkage of the binary does depend on the number of particles used in the simulations. They find a mild enhancement between their spherical and non-spherical models, of less than two, which translates into a warning in the extrapolation of numerical simulations to real galaxies \citep{Vasiliev:2013kl}. While it is true that it is very unlikely that the MW has recently  had a major merger, a minor merger is not ruled out. Indeed, as we can see, for example,  in Figure\,21 of \citet{Genzel:2010fk}, there is a significant part of the parameter space that still allows the existence of intermediate-mass MBHs in the GC (although  there is a lack of evidence of any such object).  In a broader context, if the MWNSC deviates from spherical symmetry, the same can be true for other nuclei, which might be harbouring binaries of SMBHs.

From a theoretical standpoint, both tidal disruption or gravitational capture event rates and the last parsec problem in gas-poor galaxies remain open, and the input from observational data is crucial in our understanding of these scenarios.

\section{Summary}

We have analysed Spitzer/IRAC images to measure the overall properties of the Milky Way's nuclear star cluster in a $4.5\,\mu$m image that was corrected for PAH-emission and extinction. We show that the MWNSC is not spherically symmetric, but appears to be point-symmetric in projection. We found that the cluster is strongly flattened and aligned along the Galactic plane, as has been found for nuclear star clusters in other edge-on galaxies. According to our measurements, the Milky Way's nuclear star cluster is centred on the massive black hole, Sagittarius\,A*, and has an axis ratio of $0.71\pm0.02$, a half-light radius of $(4.15\pm0.35)$\,pc, and a total luminosity and mass of $L_{NSC,4.5\,\mu{m}}=4.1\pm0.4\times10^{7}\,L_{\odot}$ and $M_{MWNSC}=2.5\pm0.4\times10^{7}$\,M$_{\odot}$, respectively. 

The size and mass of the MWNSC agree well with the corresponding values for extragalactic NSCs. The flattening of the MWNSC, along with its previously found rotation parallel to overall Galactic rotation, support the idea that it has accumulated its mass by infall of stars and gas from directions preferably along the Galactic plane and can thus be considered to ``know'' about the existence of the Galactic disc.

Models of the kinematics and evolution of the MWNSC have so far generally assumed spherical symmetry and, frequently, no rotation of the cluster. Observations show that both assumptions contradict the global properties of the NSCs at the centre of the Milky Way and of other galaxies. Future simulations of stellar dynamics around MBHs will have to deal with the considerable complexity of NSCs unveiled by high-angular NIR imaging and spectroscopy in the past decade.

\begin{acknowledgements}
 RS acknowledges support by grants AYA2009-13036, AYA2010-17631 and AYA2012-38491-CO2-02, cofunded with FEDER funds, by the Ram\' on y Cajal programme of the Spanish Ministry of Economy and Competitiveness, and by the ESO Scientific Visitor Programme. The research leading to these results has received funding from the European Research Council under the European Union's Seventh Framework Programme (FP/2007-2013) / ERC Grant Agreement n. [614922]. We thank Rick Arendt for his work on the IRAC/Spitzer maps. DK acknowledges support by the Czech Science Foundation - Deutsche Forschungsgemeinschaft collaboration project 13-00070J.
\end{acknowledgements}

\bibliography{/Users/rainer/Documents/BibDesk/BibGC}

\begin{thebibliography}{95}
\expandafter\ifx\csname natexlab\endcsname\relax\def\natexlab#1{#1}\fi

\bibitem[{{Alexander}(2006)}]{Alexander:2006fk}
{Alexander}, T. 2006, Journal of Physics Conference Series, 54, 243

\bibitem[{{Amaro-Seoane}(2012)}]{Amaro-Seoane:2012nx}
{Amaro-Seoane}, P. 2012, ArXiv e-prints

\bibitem[{{Amaro-Seoane} \& {Chen}(2014)}]{Amaro-Seoane:2014fk}
{Amaro-Seoane}, P. \& {Chen}, X. 2014, \apjl, 781, L18

\bibitem[{{Antonini}(2013)}]{Antonini:2013ys}
{Antonini}, F. 2013, \apj, 763, 62

\bibitem[{{Antonini} {et~al.}(2012){Antonini}, {Capuzzo-Dolcetta},
  {Mastrobuono-Battisti}, \& {Merritt}}]{Antonini:2012fk}
{Antonini}, F., {Capuzzo-Dolcetta}, R., {Mastrobuono-Battisti}, A., \&
  {Merritt}, D. 2012, \apj, 750, 111

\bibitem[{{Arendt} {et~al.}(2008){Arendt}, {Stolovy}, {Ram{\'{\i}}rez},
  {Sellgren}, {Cotera}, {Law}, {Yusef-Zadeh}, {Smith}, \&
  {Gezari}}]{Arendt:2008fk}
{Arendt}, R.~G., {Stolovy}, S.~R., {Ram{\'{\i}}rez}, S.~V., {et~al.} 2008,
  \apj, 682, 384

\bibitem[{{Bahcall} \& {Wolf}(1977)}]{Bahcall:1977ys}
{Bahcall}, J.~N. \& {Wolf}, R.~A. 1977, \apj, 216, 883

\bibitem[{{Bartko} {et~al.}(2009){Bartko}, {Martins}, {Fritz}, {Genzel},
  {Levin}, {Perets}, {Paumard}, {Nayakshin}, {Gerhard}, {Alexander},
  {Dodds-Eden}, {Eisenhauer}, {Gillessen}, {Mascetti}, {Ott}, {Perrin},
  {Pfuhl}, {Reid}, {Rouan}, {Sternberg}, \& {Trippe}}]{Bartko:2009fq}
{Bartko}, H., {Martins}, F., {Fritz}, T.~K., {et~al.} 2009, \apj, 697, 1741

\bibitem[{{Bartko} {et~al.}(2010){Bartko}, {Martins}, {Trippe}, {Fritz},
  {Genzel}, {Ott}, {Eisenhauer}, {Gillessen}, {Paumard}, {Alexander},
  {Dodds-Eden}, {Gerhard}, {Levin}, {Mascetti}, {Nayakshin}, {Perets},
  {Perrin}, {Pfuhl}, {Reid}, {Rouan}, {Zilka}, \& {Sternberg}}]{Bartko:2010fk}
{Bartko}, H., {Martins}, F., {Trippe}, S., {et~al.} 2010, \apj, 708, 834

\bibitem[{{Becklin} \& {Neugebauer}(1968)}]{Becklin:1968nx}
{Becklin}, E.~E. \& {Neugebauer}, G. 1968, \apj, 151, 145

\bibitem[{{Berczik} {et~al.}(2006){Berczik}, {Merritt}, {Spurzem}, \&
  {Bischof}}]{Berczik:2006kx}
{Berczik}, P., {Merritt}, D., {Spurzem}, R., \& {Bischof}, H.-P. 2006, \apjl,
  642, L21

\bibitem[{{B{\"o}ker}(2010)}]{Boker:2010ys}
{B{\"o}ker}, T. 2010, in IAU Symposium, Vol. 266, IAU Symposium, ed. {R.~de
  Grijs \& J.~R.~D.~L{\'e}pine}, 58--63

\bibitem[{{B{\"o}ker} {et~al.}(2002){B{\"o}ker}, {Laine}, {van der Marel},
  {Sarzi}, {Rix}, {Ho}, \& {Shields}}]{Boker:2002kx}
{B{\"o}ker}, T., {Laine}, S., {van der Marel}, R.~P., {et~al.} 2002, \aj, 123,
  1389

\bibitem[{{B{\"o}ker} {et~al.}(2004){B{\"o}ker}, {Sarzi}, {McLaughlin}, {van
  der Marel}, {Rix}, {Ho}, \& {Shields}}]{Boker:2004oq}
{B{\"o}ker}, T., {Sarzi}, M., {McLaughlin}, D.~E., {et~al.} 2004, \aj, 127, 105

\bibitem[{{Buchholz} {et~al.}(2009){Buchholz}, {Sch{\"o}del}, \&
  {Eckart}}]{Buchholz:2009fk}
{Buchholz}, R.~M., {Sch{\"o}del}, R., \& {Eckart}, A. 2009, \aap, 499, 483

\bibitem[{{Cappellari}(2002)}]{Cappellari:2002qf}
{Cappellari}, M. 2002, \mnras, 333, 400

\bibitem[{{Carollo} {et~al.}(1998){Carollo}, {Stiavelli}, \&
  {Mack}}]{Carollo:1998fk}
{Carollo}, C.~M., {Stiavelli}, M., \& {Mack}, J. 1998, \aj, 116, 68

\bibitem[{{Catchpole} {et~al.}(1990){Catchpole}, {Whitelock}, \&
  {Glass}}]{Catchpole:1990rz}
{Catchpole}, R.~M., {Whitelock}, P.~A., \& {Glass}, I.~S. 1990, \mnras, 247,
  479

\bibitem[{{C{\^o}t{\'e}} {et~al.}(2006){C{\^o}t{\'e}}, {Piatek}, {Ferrarese},
  {Jord{\'a}n}, {Merritt}, {Peng}, {Ha{\c s}egan}, {Blakeslee}, {Mei}, {West},
  {Milosavljevi{\'c}}, \& {Tonry}}]{Cote:2006eu}
{C{\^o}t{\'e}}, P., {Piatek}, S., {Ferrarese}, L., {et~al.} 2006, \apjs, 165,
  57

\bibitem[{{Crocker} {et~al.}(2010){Crocker}, {Jones}, {Melia}, {Ott}, \&
  {Protheroe}}]{Crocker:2010fk}
{Crocker}, R.~M., {Jones}, D.~I., {Melia}, F., {Ott}, J., \& {Protheroe}, R.~J.
  2010, \nat, 463, 65

\bibitem[{{Dale} {et~al.}(2009){Dale}, {Davies}, {Church}, \&
  {Freitag}}]{Dale:2009ca}
{Dale}, J.~E., {Davies}, M.~B., {Church}, R.~P., \& {Freitag}, M. 2009, \mnras,
  393, 1016

\bibitem[{{Diolaiti} {et~al.}(2000){Diolaiti}, {Bendinelli}, {Bonaccini},
  {Close}, {Currie}, \& {Parmeggiani}}]{Diolaiti:2000qo}
{Diolaiti}, E., {Bendinelli}, O., {Bonaccini}, D., {et~al.} 2000, \aaps, 147,
  335

\bibitem[{{Do} {et~al.}(2009){Do}, {Ghez}, {Morris}, {Lu}, {Matthews}, {Yelda},
  \& {Larkin}}]{Do:2009tg}
{Do}, T., {Ghez}, A.~M., {Morris}, M.~R., {et~al.} 2009, \apj, 703, 1323

\bibitem[{{Do} {et~al.}(2013){Do}, {Lu}, {Ghez}, {Morris}, {Yelda}, {Martinez},
  {Wright}, \& {Matthews}}]{Do:2013fk}
{Do}, T., {Lu}, J.~R., {Ghez}, A.~M., {et~al.} 2013, \apj, 764, 154

\bibitem[{{Eckart} {et~al.}(1993){Eckart}, {Genzel}, {Hofmann}, {Sams}, \&
  {Tacconi-Garman}}]{Eckart:1993zr}
{Eckart}, A., {Genzel}, R., {Hofmann}, R., {Sams}, B.~J., \& {Tacconi-Garman},
  L.~E. 1993, \apjl, 407, L77

\bibitem[{{Emsellem} {et~al.}(1994){Emsellem}, {Monnet}, \&
  {Bacon}}]{Emsellem:1994ve}
{Emsellem}, E., {Monnet}, G., \& {Bacon}, R. 1994, \aap, 285, 723

\bibitem[{{Fritz} {et~al.}(2013){Fritz}, {Chatzopoulos}, {Gerhard},
  {Gillessen}, {Genzel}, {Pfuhl}, {Tacchella}, {Eisenhauer}, \&
  {Ott}}]{Fritz:2013hc}
{Fritz}, T.~K., {Chatzopoulos}, S., {Gerhard}, O., {et~al.} 2013, ArXiv
  e-prints

\bibitem[{{Fritz} {et~al.}(2011){Fritz}, {Gillessen}, {Dodds-Eden}, {Lutz},
  {Genzel}, {Raab}, {Ott}, {Pfuhl}, {Eisenhauer}, \&
  {Yusef-Zadeh}}]{Fritz:2011fk}
{Fritz}, T.~K., {Gillessen}, S., {Dodds-Eden}, K., {et~al.} 2011, \apj, 737, 73

\bibitem[{{Garc{\'{\i}}a-Mar{\'{\i}}n}
  {et~al.}(2011){Garc{\'{\i}}a-Mar{\'{\i}}n}, {Eckart}, {Weiss}, {Witzel},
  {Bremer}, {Zamaninasab}, {Morris}, {Sch{\"o}del}, {Kunneriath}, {Nishiyama},
  {Baganoff}, {Dov{\v c}iak}, {Sabha}, {Duschl}, {Moultaka}, {Karas},
  {Najarro}, {Mu{\v z}i{\'c}}, {Straubmeier}, {Vogel}, {Krips}, \&
  {Wiesemeyer}}]{Garcia-Marin:2011fk}
{Garc{\'{\i}}a-Mar{\'{\i}}n}, M., {Eckart}, A., {Weiss}, A., {et~al.} 2011,
  \apj, 738, 158

\bibitem[{{Genzel} {et~al.}(2010){Genzel}, {Eisenhauer}, \&
  {Gillessen}}]{Genzel:2010fk}
{Genzel}, R., {Eisenhauer}, F., \& {Gillessen}, S. 2010, Reviews of Modern
  Physics, 82, 3121

\bibitem[{{Genzel} {et~al.}(2003){Genzel}, {Sch{\"o}del}, {Ott}, {Eisenhauer},
  {Hofmann}, {Lehnert}, {Eckart}, {Alexander}, {Sternberg}, {Lenzen},
  {Cl{\'e}net}, {Lacombe}, {Rouan}, {Renzini}, \&
  {Tacconi-Garman}}]{Genzel:2003it}
{Genzel}, R., {Sch{\"o}del}, R., {Ott}, T., {et~al.} 2003, \apj, 594, 812

\bibitem[{{Ghez} {et~al.}(2009){Ghez}, {Morris}, {Lu}, {Weinberg}, {Matthews},
  {Alexander}, {Armitage}, {Becklin}, {Brown}, {Campbell}, {Do}, {Eckart},
  {Genzel}, {Gould}, {Hansen}, {Ho}, {Lo}, {Loeb}, {Melia}, {Merritt},
  {Milosavljevic}, {Perets}, {Rasio}, {Reid}, {Salim}, {Sch{\"o}del}, \&
  {Yelda}}]{Ghez:2009kx}
{Ghez}, A., {Morris}, M., {Lu}, J., {et~al.} 2009, in Astronomy, Vol. 2010,
  astro2010: The Astronomy and Astrophysics Decadal Survey, 89--+

\bibitem[{{Ghez} {et~al.}(1998){Ghez}, {Klein}, {Morris}, \&
  {Becklin}}]{Ghez:1998ad}
{Ghez}, A.~M., {Klein}, B.~L., {Morris}, M., \& {Becklin}, E.~E. 1998, \apj,
  509, 678

\bibitem[{{Ghez} {et~al.}(2008){Ghez}, {Salim}, {Weinberg}, {Lu}, {Do}, {Dunn},
  {Matthews}, {Morris}, {Yelda}, {Becklin}, {Kremenek}, {Milosavljevic}, \&
  {Naiman}}]{Ghez:2008fk}
{Ghez}, A.~M., {Salim}, S., {Weinberg}, N.~N., {et~al.} 2008, \apj, 689, 1044

\bibitem[{{Gillessen} {et~al.}(2009){Gillessen}, {Eisenhauer}, {Trippe},
  {Alexander}, {Genzel}, {Martins}, \& {Ott}}]{Gillessen:2009qe}
{Gillessen}, S., {Eisenhauer}, F., {Trippe}, S., {et~al.} 2009, \apj, 692, 1075

\bibitem[{{Gnedin} {et~al.}(2013){Gnedin}, {Ostriker}, \&
  {Tremaine}}]{Gnedin:2013vn}
{Gnedin}, O.~Y., {Ostriker}, J.~P., \& {Tremaine}, S. 2013, ArXiv e-prints

\bibitem[{{Graham}(2001)}]{Graham:2001ys}
{Graham}, A.~W. 2001, \aj, 121, 820

\bibitem[{{Graham} \& {Spitler}(2009)}]{Graham:2009lh}
{Graham}, A.~W. \& {Spitler}, L.~R. 2009, \mnras, 397, 1003

\bibitem[{{Hartmann} {et~al.}(2011){Hartmann}, {Debattista}, {Seth},
  {Cappellari}, \& {Quinn}}]{Hartmann:2011uq}
{Hartmann}, M., {Debattista}, V.~P., {Seth}, A., {Cappellari}, M., \& {Quinn},
  T.~R. 2011, \mnras, 418, 2697

\bibitem[{{Holley-Bockelmann} {et~al.}(2001){Holley-Bockelmann}, {Mihos},
  {Sigurdsson}, \& {Hernquist}}]{Holley-Bockelmann:2001fk}
{Holley-Bockelmann}, K., {Mihos}, J.~C., {Sigurdsson}, S., \& {Hernquist}, L.
  2001, \apj, 549, 862

\bibitem[{{Holley-Bockelmann} {et~al.}(2002){Holley-Bockelmann}, {Mihos},
  {Sigurdsson}, {Hernquist}, \& {Norman}}]{Holley-Bockelmann:2002uq}
{Holley-Bockelmann}, K., {Mihos}, J.~C., {Sigurdsson}, S., {Hernquist}, L., \&
  {Norman}, C. 2002, \apj, 567, 817

\bibitem[{{Khan} {et~al.}(2013){Khan}, {Holley-Bockelmann}, {Berczik}, \&
  {Just}}]{Khan:2013zr}
{Khan}, F.~M., {Holley-Bockelmann}, K., {Berczik}, P., \& {Just}, A. 2013,
  \apj, 773, 100

\bibitem[{{King}(1962)}]{King:1962kx}
{King}, I. 1962, \aj, 67, 471

\bibitem[{{Krabbe} {et~al.}(1995){Krabbe}, {Genzel}, {Eckart}, {Najarro},
  {Lutz}, {Cameron}, {Kroker}, {Tacconi-Garman}, {Thatte}, {Weitzel},
  {Drapatz}, {Geballe}, {Sternberg}, \& {Kudritzki}}]{Krabbe:1995fk}
{Krabbe}, A., {Genzel}, R., {Eckart}, A., {et~al.} 1995, \apjl, 447, L95+

\bibitem[{{Kurucz}(1993)}]{Kurucz:1993fk}
{Kurucz}, R.~L. 1993, VizieR Online Data Catalog, 6039, 0

\bibitem[{{Launhardt} {et~al.}(2002){Launhardt}, {Zylka}, \&
  {Mezger}}]{Launhardt:2002nx}
{Launhardt}, R., {Zylka}, R., \& {Mezger}, P.~G. 2002, \aap, 384, 112

\bibitem[{{Levin} \& {Beloborodov}(2003)}]{Levin:2003kx}
{Levin}, Y. \& {Beloborodov}, A.~M. 2003, \apjl, 590, L33

\bibitem[{{Lightman} \& {Shapiro}(1977)}]{Lightman:1977ly}
{Lightman}, A.~P. \& {Shapiro}, S.~L. 1977, \apj, 211, 244

\bibitem[{{Longmore} {et~al.}(2012){Longmore}, {Rathborne}, {Bastian}, {Alves},
  {Ascenso}, {Bally}, {Testi}, {Longmore}, {Battersby}, {Bressert}, {Purcell},
  {Walsh}, {Jackson}, {Foster}, {Molinari}, {Meingast}, {Amorim}, {Lima},
  {Marques}, {Moitinho}, {Pinhao}, {Rebordao}, \& {Santos}}]{Longmore:2012uq}
{Longmore}, S.~N., {Rathborne}, J., {Bastian}, N., {et~al.} 2012, \apj, 746,
  117

\bibitem[{{Lu} {et~al.}(2013){Lu}, {Do}, {Ghez}, {Morris}, {Yelda}, \&
  {Matthews}}]{Lu:2013fk}
{Lu}, J.~R., {Do}, T., {Ghez}, A.~M., {et~al.} 2013, \apj, 764, 155

\bibitem[{{Lu} {et~al.}(2009){Lu}, {Ghez}, {Hornstein}, {Morris}, {Becklin}, \&
  {Matthews}}]{Lu:2009bl}
{Lu}, J.~R., {Ghez}, A.~M., {Hornstein}, S.~D., {et~al.} 2009, \apj, 690, 1463

\bibitem[{{Majewski} {et~al.}(2011){Majewski}, {Zasowski}, \&
  {Nidever}}]{Majewski:2011uq}
{Majewski}, S.~R., {Zasowski}, G., \& {Nidever}, D.~L. 2011, \apj, 739, 25

\bibitem[{{Malkin}(2012)}]{Malkin:2012uq}
{Malkin}, Z. 2012, ArXiv e-prints

\bibitem[{{Markwardt}(2009)}]{Markwardt:2009fk}
{Markwardt}, C.~B. 2009, in Astronomical Society of the Pacific Conference
  Series, Vol. 411, Astronomical Data Analysis Software and Systems XVIII, ed.
  D.~A. {Bohlender}, D.~{Durand}, \& P.~{Dowler}, 251

\bibitem[{{Meidt} {et~al.}(2014){Meidt}, {Schinnerer}, {van de Ven},
  {Zaritsky}, {Peletier}, {Knapen}, {Sheth}, {Regan}, {Querejeta},
  {Munoz-Mateos}, {Kim}, {Hinz}, {Gil de Paz}, {Athanassoula}, {Bosma}, {Buta},
  {Cisternas}, {Ho}, {Holwerda}, {Skibba}, {Laurikainen}, {Salo}, {Gadotti},
  {Laine}, {Erroz-Ferrer}, {Comeron}, {Menendez-Delmestre}, {Seibert}, \&
  {Mizusawa}}]{Meidt:2014fk}
{Meidt}, S.~E., {Schinnerer}, E., {van de Ven}, G., {et~al.} 2014, ArXiv
  e-prints

\bibitem[{{Merritt}(2010)}]{Merritt:2010ve}
{Merritt}, D. 2010, \apj, 718, 739

\bibitem[{{Merritt}(2013)}]{Merritt:2013uq}
{Merritt}, D. 2013, {Dynamics and Evolution of Galactic Nuclei} (Princeton
  University Press)

\bibitem[{{Merritt} \& {Poon}(2004)}]{Merritt:2004bh}
{Merritt}, D. \& {Poon}, M.~Y. 2004, \apj, 606, 788

\bibitem[{{Merritt} \& {Vasiliev}(2011)}]{Merritt:2011cr}
{Merritt}, D. \& {Vasiliev}, E. 2011, \apj, 726, 61

\bibitem[{{Milosavljevi{\'c}} \& {Merritt}(2001)}]{Milosavljevic:2001ly}
{Milosavljevi{\'c}}, M. \& {Merritt}, D. 2001, \apj, 563, 34

\bibitem[{{Misgeld} \& {Hilker}(2011)}]{Misgeld:2011kx}
{Misgeld}, I. \& {Hilker}, M. 2011, \mnras, 414, 3699

\bibitem[{{Moffat}(1969)}]{Moffat:1969zr}
{Moffat}, A.~F.~J. 1969, \aap, 3, 455

\bibitem[{{Molinari} {et~al.}(2011){Molinari}, {Bally}, {Noriega-Crespo},
  {Compi{\`e}gne}, {Bernard}, {Paradis}, {Martin}, {Testi}, {Barlow}, {Moore},
  {Plume}, {Swinyard}, {Zavagno}, {Calzoletti}, {Di Giorgio}, {Elia},
  {Faustini}, {Natoli}, {Pestalozzi}, {Pezzuto}, {Piacentini}, {Polenta},
  {Polychroni}, {Schisano}, {Traficante}, {Veneziani}, {Battersby}, {Burton},
  {Carey}, {Fukui}, {Li}, {Lord}, {Morgan}, {Motte}, {Schuller},
  {Stringfellow}, {Tan}, {Thompson}, {Ward-Thompson}, {White}, \&
  {Umana}}]{Molinari:2011fk}
{Molinari}, S., {Bally}, J., {Noriega-Crespo}, A., {et~al.} 2011, \apjl, 735,
  L33

\bibitem[{{Morris} \& {Serabyn}(1996)}]{Morris:1996vn}
{Morris}, M. \& {Serabyn}, E. 1996, \araa, 34, 645

\bibitem[{{Murphy} {et~al.}(1991){Murphy}, {Cohn}, \&
  {Durisen}}]{Murphy:1991zr}
{Murphy}, B.~W., {Cohn}, H.~N., \& {Durisen}, R.~H. 1991, \apj, 370, 60

\bibitem[{{Neumayer} \& {Walcher}(2012)}]{Neumayer:2012fk}
{Neumayer}, N. \& {Walcher}, C.~J. 2012, Advances in Astronomy, 2012

\bibitem[{{Neumayer} {et~al.}(2011){Neumayer}, {Walcher}, {Andersen},
  {S{\'a}nchez}, {B{\"o}ker}, \& {Rix}}]{Neumayer:2011uq}
{Neumayer}, N., {Walcher}, C.~J., {Andersen}, D., {et~al.} 2011, \mnras, 413,
  1875

\bibitem[{{Nishiyama} {et~al.}(2009){Nishiyama}, {Tamura}, {Hatano}, {Kato},
  {Tanab{\'e}}, {Sugitani}, \& {Nagata}}]{Nishiyama:2009oj}
{Nishiyama}, S., {Tamura}, M., {Hatano}, H., {et~al.} 2009, \apj, 696, 1407

\bibitem[{{Oh} {et~al.}(2008){Oh}, {de Blok}, {Walter}, {Brinks}, \&
  {Kennicutt}}]{Oh:2008fk}
{Oh}, S.-H., {de Blok}, W.~J.~G., {Walter}, F., {Brinks}, E., \& {Kennicutt},
  Jr., R.~C. 2008, \aj, 136, 2761

\bibitem[{{Paumard} {et~al.}(2006){Paumard}, {Genzel}, {Martins}, {Nayakshin},
  {Beloborodov}, {Levin}, {Trippe}, {Eisenhauer}, {Ott}, {Gillessen}, {Abuter},
  {Cuadra}, {Alexander}, \& {Sternberg}}]{Paumard:2006xd}
{Paumard}, T., {Genzel}, R., {Martins}, F., {et~al.} 2006, \apj, 643, 1011

\bibitem[{{Pfuhl} {et~al.}(2011){Pfuhl}, {Fritz}, {Zilka}, {Maness},
  {Eisenhauer}, {Genzel}, {Gillessen}, {Ott}, {Dodds-Eden}, \&
  {Sternberg}}]{Pfuhl:2011uq}
{Pfuhl}, O., {Fritz}, T.~K., {Zilka}, M., {et~al.} 2011, \apj, 741, 108

\bibitem[{{Philipp} {et~al.}(1999){Philipp}, {Zylka}, {Mezger}, {Duschl},
  {Herbst}, \& {Tuffs}}]{Philipp:1999nx}
{Philipp}, S., {Zylka}, R., {Mezger}, P.~G., {et~al.} 1999, \aap, 348, 768

\bibitem[{{Poon} \& {Merritt}(2001)}]{Poon:2001qf}
{Poon}, M.~Y. \& {Merritt}, D. 2001, \apj, 549, 192

\bibitem[{{Poon} \& {Merritt}(2004)}]{Poon:2004dq}
{Poon}, M.~Y. \& {Merritt}, D. 2004, \apj, 606, 774

\bibitem[{{Portegies Zwart} {et~al.}(2002){Portegies Zwart}, {Makino},
  {McMillan}, \& {Hut}}]{Portegies-Zwart:2002fk}
{Portegies Zwart}, S.~F., {Makino}, J., {McMillan}, S.~L.~W., \& {Hut}, P.
  2002, \apj, 565, 265

\bibitem[{{Preto} \& {Amaro-Seoane}(2010)}]{Preto:2010kx}
{Preto}, M. \& {Amaro-Seoane}, P. 2010, \apjl, 708, L42

\bibitem[{{Ram{\'{\i}}rez} {et~al.}(2008){Ram{\'{\i}}rez}, {Arendt},
  {Sellgren}, {Stolovy}, {Cotera}, {Smith}, \& {Yusef-Zadeh}}]{Ramirez:2008fk}
{Ram{\'{\i}}rez}, S.~V., {Arendt}, R.~G., {Sellgren}, K., {et~al.} 2008, \apjs,
  175, 147

\bibitem[{{Rauch}(1995)}]{Rauch:1995ve}
{Rauch}, K.~P. 1995, \mnras, 275, 628

\bibitem[{{Rossa} {et~al.}(2006){Rossa}, {van der Marel}, {B{\"o}ker},
  {Gerssen}, {Ho}, {Rix}, {Shields}, \& {Walcher}}]{Rossa:2006zr}
{Rossa}, J., {van der Marel}, R.~P., {B{\"o}ker}, T., {et~al.} 2006, \aj, 132,
  1074

\bibitem[{{Sch{\"o}del}(2011)}]{Schodel:2011ab}
{Sch{\"o}del}, R. 2011, in Astronomical Society of the Pacific Conference
  Series, Vol. 439, Astronomical Society of the Pacific Conference Series, ed.
  {M.~R.~Morris, Q.~D.~Wang, \& F.~Yuan}, 222--+

\bibitem[{{Sch{\"o}del} {et~al.}(2007){Sch{\"o}del}, {Eckart}, {Alexander},
  {Merritt}, {Genzel}, {Sternberg}, {Meyer}, {Kul}, {Moultaka}, {Ott}, \&
  {Straubmeier}}]{Schodel:2007tw}
{Sch{\"o}del}, R., {Eckart}, A., {Alexander}, T., {et~al.} 2007, \aap, 469, 125

\bibitem[{{Sch{\"o}del} {et~al.}(2009){Sch{\"o}del}, {Merritt}, \&
  {Eckart}}]{Schodel:2009zr}
{Sch{\"o}del}, R., {Merritt}, D., \& {Eckart}, A. 2009, \aap, 502, 91

\bibitem[{{Sch{\"o}del} {et~al.}(2010){Sch{\"o}del}, {Najarro}, {Muzic}, \&
  {Eckart}}]{Schodel:2010fk}
{Sch{\"o}del}, R., {Najarro}, F., {Muzic}, K., \& {Eckart}, A. 2010, \aap, 511,
  A18+

\bibitem[{{Scoville} {et~al.}(2003){Scoville}, {Stolovy}, {Rieke},
  {Christopher}, \& {Yusef-Zadeh}}]{Scoville:2003la}
{Scoville}, N.~Z., {Stolovy}, S.~R., {Rieke}, M., {Christopher}, M., \&
  {Yusef-Zadeh}, F. 2003, \apj, 594, 294

\bibitem[{{Seth} {et~al.}(2008{\natexlab{a}}){Seth}, {Ag{\"u}eros}, {Lee}, \&
  {Basu-Zych}}]{Seth:2008rr}
{Seth}, A., {Ag{\"u}eros}, M., {Lee}, D., \& {Basu-Zych}, A.
  2008{\natexlab{a}}, \apj, 678, 116

\bibitem[{{Seth} {et~al.}(2008{\natexlab{b}}){Seth}, {Blum}, {Bastian},
  {Caldwell}, \& {Debattista}}]{Seth:2008kx}
{Seth}, A.~C., {Blum}, R.~D., {Bastian}, N., {Caldwell}, N., \& {Debattista},
  V.~P. 2008{\natexlab{b}}, \apj, 687, 997

\bibitem[{{Seth} {et~al.}(2010){Seth}, {Cappellari}, {Neumayer}, {Caldwell},
  {Bastian}, {Olsen}, {Blum}, {Debattista}, {McDermid}, {Puzia}, \&
  {Stephens}}]{Seth:2010fk}
{Seth}, A.~C., {Cappellari}, M., {Neumayer}, N., {et~al.} 2010, \apj, 714, 713

\bibitem[{{Seth} {et~al.}(2006){Seth}, {Dalcanton}, {Hodge}, \&
  {Debattista}}]{Seth:2006uq}
{Seth}, A.~C., {Dalcanton}, J.~J., {Hodge}, P.~W., \& {Debattista}, V.~P. 2006,
  \aj, 132, 2539

\bibitem[{{Stolovy} {et~al.}(2006){Stolovy}, {Ramirez}, {Arendt}, {Cotera},
  {Yusef-Zadeh}, {Law}, {Gezari}, {Sellgren}, {Karr}, {Moseley}, \&
  {Smith}}]{Stolovy:2006fk}
{Stolovy}, S., {Ramirez}, S., {Arendt}, R.~G., {et~al.} 2006, Journal of
  Physics Conference Series, 54, 176

\bibitem[{{Trippe} {et~al.}(2008){Trippe}, {Gillessen}, {Gerhard}, {Bartko},
  {Fritz}, {Maness}, {Eisenhauer}, {Martins}, {Ott}, {Dodds-Eden}, \&
  {Genzel}}]{Trippe:2008it}
{Trippe}, S., {Gillessen}, S., {Gerhard}, O.~E., {et~al.} 2008, \aap, 492, 419

\bibitem[{{Vasiliev} {et~al.}(2013){Vasiliev}, {Antonini}, \&
  {Merritt}}]{Vasiliev:2013kl}
{Vasiliev}, E., {Antonini}, F., \& {Merritt}, D. 2013, ArXiv e-prints

\bibitem[{{Vasiliev} \& {Merritt}(2013{\natexlab{a}})}]{Vasiliev:2013ys}
{Vasiliev}, E. \& {Merritt}, D. 2013{\natexlab{a}}, \apj, 774, 87

\bibitem[{{Vasiliev} \& {Merritt}(2013{\natexlab{b}})}]{Vasiliev:2013oq}
{Vasiliev}, E. \& {Merritt}, D. 2013{\natexlab{b}}, \apj, 774, 87

\bibitem[{{Walcher} {et~al.}(2006){Walcher}, {B{\"o}ker}, {Charlot}, {Ho},
  {Rix}, {Rossa}, {Shields}, \& {van der Marel}}]{Walcher:2006ve}
{Walcher}, C.~J., {B{\"o}ker}, T., {Charlot}, S., {et~al.} 2006, \apj, 649, 692

\bibitem[{{Walcher} {et~al.}(2005){Walcher}, {van der Marel}, {McLaughlin},
  {Rix}, {B{\"o}ker}, {H{\"a}ring}, {Ho}, {Sarzi}, \&
  {Shields}}]{Walcher:2005ys}
{Walcher}, C.~J., {van der Marel}, R.~P., {McLaughlin}, D., {et~al.} 2005,
  \apj, 618, 237

\end{thebibliography}

\end{document}